\documentclass[rmp]{revtex4}
\usepackage{amssymb,amsfonts,amsmath}
\usepackage{hyperref}

\usepackage{graphicx}
\usepackage{amsmath}
\usepackage{natbib}

\newcommand{\EQ}[1]{Eq.~(\ref{eq:#1})}

\newcommand{\FIG}[1]{Fig.~\ref{fig:#1}}

\newcommand{\x}{x}
\newcommand{\xs}{\bar{\x}}
\newcommand{\xz}{z}
\newcommand{\xzs}{\bar{\xz}}
\newcommand{\dx}{\delta \x}
\newcommand{\n}{n}
\newcommand{\rate}{\gamma}
\newcommand{\ns}{\bar{\n}}

\newcommand{\mk}{\bar{k}}
\newcommand{\dk}{\delta \bar{k}}
\newcommand{\tx}[1]{\xz_{#1}}
\newcommand{\mr}[1]{\psi^{(#1)}}
\newcommand{\ml}[1]{\phi^{(#1)}}
\newcommand{\Smin}{S^*}
\newcommand{\mut}{u}
\newcommand{\cp}{\iota}

\newcommand{\la}{\langle}
\newcommand{\ra}{\rangle}

\begin{document}

\title{Fluctuations of fitness distributions and the rate of Muller's ratchet}
\author{Richard~A.~Neher}
\affiliation{Max Planck Institute for Developmental Biology, T\"ubingen, 72076, Germany}
\author{Boris~I.~Shraiman}
\affiliation{Kavli Institute for Theoretical Physics and Department of Physics, University of California, Santa Barbara, CA 93106}
\date{\today}

\begin{abstract}
The accumulation of deleterious mutations is driven by rare fluctuations which lead to the loss of all mutation free individuals, a process known as Muller's ratchet. Even though Muller's ratchet is a paradigmatic process in population genetics, a quantitative understanding of its rate is still lacking. The difficulty lies in the nontrivial nature of fluctuations in the fitness distribution which control the rate of extinction of the fittest genotype. We address this problem using the simple but classic model of mutation selection balance with deleterious mutations all having the same effect on fitness. We show analytically how fluctuations among the fittest individuals propagate to individuals of lower fitness and have a dramatically amplified effects on the bulk of the population at a later time. 
If a reduction in the size of the fittest class reduces the mean fitness only after a delay, selection opposing this reduction is also delayed. This delayed restoring force speeds up Muller's ratchet. We show how the delayed response can be accounted for using a path integral formulation of the stochastic dynamics and provide an expression for the rate of the ratchet that is accurate across a broad range of parameters.
\end{abstract}
\maketitle
By weeding out deleterious mutations, purifying selection acts to preserve a functional genome. In sufficiently small populations, however, weakly deleterious mutations can by chance fix. This phenomenon, termed Muller's ratchet \citep{Muller:1964p45018,Felsenstein:1974p23937}, is especially important in the absence of recombination and is thought to account for the degeneration of Y-chromosomes \citep{Rice:1987p45218} and for the absence of long lived asexual lineages \citep{Lynch:1993p42844}. 

\begin{figure}[b]
\begin{center}
  \includegraphics[width=0.5\columnwidth]{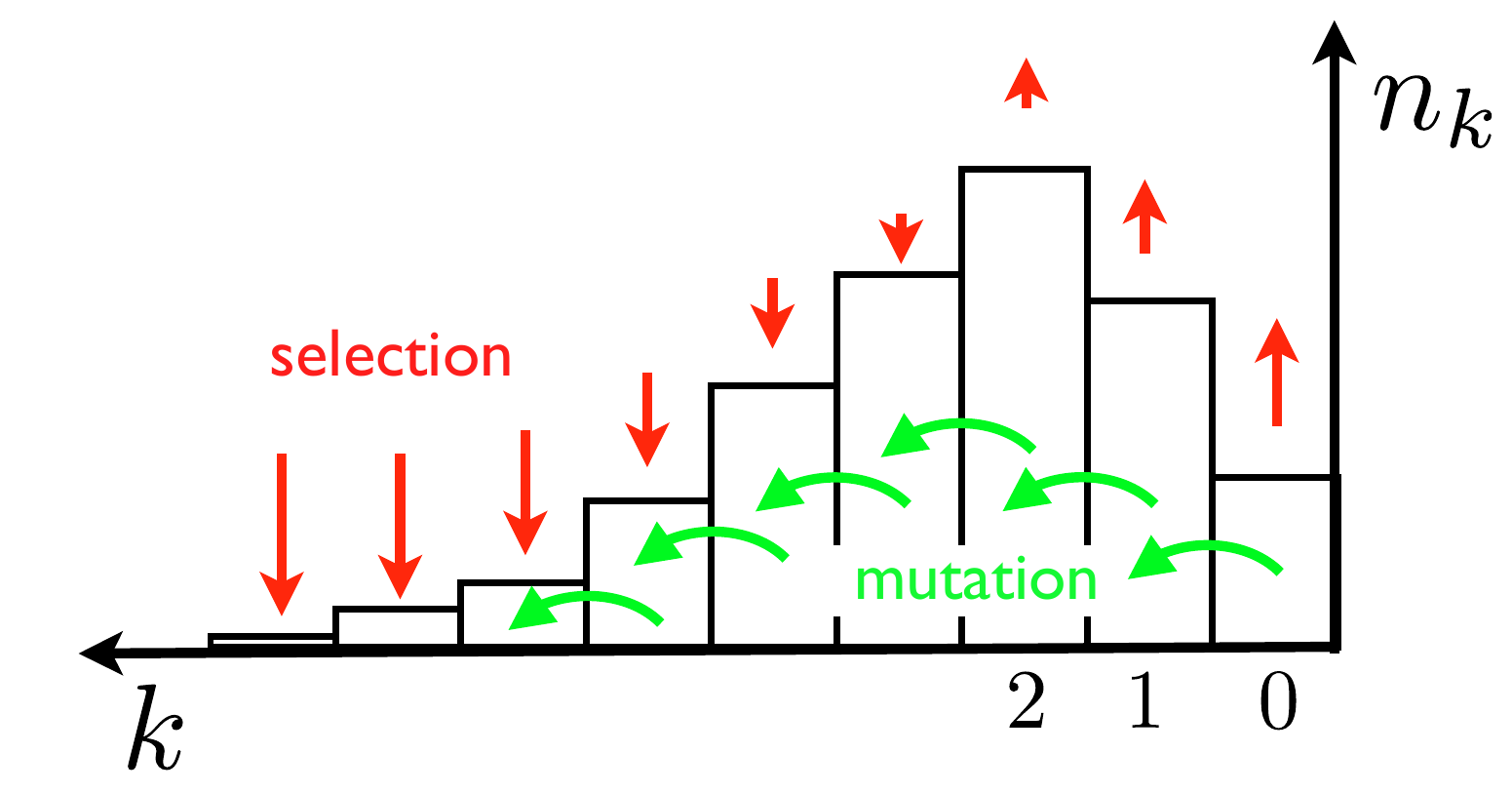}
  \caption[labelInTOC]{Deleterious mutation selection balance. The population is distributed among classes of individuals carrying $k$ deleterious mutations. Classes with few mutations grow due to selection (red arrows), but loose individuals through mutations (green arrows), while classes with many mutations are selected against but replenished by mutations. }
  \label{fig:sketch}
\end{center}
\end{figure}

A click of Muller's ratchet refers to the loss of the class of individuals with the smallest number of deleterious mutations. To understand the processes responsible for such a click, it is useful to consider a simple model of accumulation of deleterious mutations with identical effect sizes illustrated in \FIG{sketch}. Because of mutations, the population spreads out along the fitness axis, which in this model is equivalent to the number of deleterious mutations in a genome. The population can hence be grouped into discrete classes each characterized by the number of deleterious mutations. Mutation carries individuals from classes with fewer to classes with more mutations, hence shifting the population to the left. This tendency is opposed by selection, which amplifies fit individuals on the right, while decreasing the number of unfit individuals on the left. These opposing trends lead to a steady balance, at least in sufficiently large populations. However, in addition to selection and mutation, the distribution of individuals among fitness classes is affected by fluctuations in the number of offspring produced by individuals of different classes, i.e., by genetic drift. Such fluctuations are stronger (in relative terms) in smaller populations and in particular in classes which carry only a small number of individuals.  When mutation rate is high and selection is weak, the class of individuals with the smallest number of mutations ($k=0$ in \FIG{sketch}) contains only few individuals and is therefore susceptible to accidental extinction - an event that corresponds to the ``click" of Muller's ratchet.

Despite the simplicity of the classic model described above, understanding the rate of the ratchet has been a challenge and remains incomplete \citep{Stephan:1993p42929,Higgs:1995p45226,Gessler:1995p42788,Gordo:2000p42688,Stephan:2002_review,Etheridge:2007p44291,Jain:2008p45047,Waxman:2010p47020}. Here, we revisit this problem starting with the systematic analysis of  fluctuations in the distribution of the population among different fitness classes. We show that fitness classes do not fluctuate independently. Instead, there are collective modes affecting the entire distribution which relax on different time scales. Having identified these modes, we calculate the fluctuations of the number of individuals in the fittest class and show how these fluctuations affect the mean fitness. Fluctuations in mean fitness feed back on the fittest class with a delay and thereby control the probability of extinction. These insights allow us to arrive at a better approximation to the rate of the ratchet. In particular, we show that the parameter introduced in the earlier work \citep{Haigh:1978p37141,Stephan:1993p42929,Gordo:2000p42688} to parameterize the effective strength of selection in the least loaded class is not a constant but depends on the ratio of the mutation rate and the effect size of mutations. We use the path integral representation of stochastic processes, borrowed from physics \citep{Feynman:1965} to describe the dynamics of the fittest class and arrive at an approximation of the rate of Muller's ratchet that is accurate across a large parameter range.

Understanding the rate of the ratchet is important, for example to estimate the number of beneficial mutations required to halt the ratchet and prevent the mutational melt-down of a population \citep{Lynch:1993p42844,Pfaffelhuber:2011p44301,Goyal:2012p47382} (for an in-depth and up-to-date discussion of the importance of deleterious mutations we refer the reader to \citet{Charlesworth:2012p45100}). Furthermore, fluctuations of fitness distributions are a general phenomenon with profound implications for the dynamics of adaptation and genetic diversity of populations. Below we shall place our approach into the context of the recent studies of the dynamics of adaptation in populations with extensive non-neutral genetic diversity \citep{Tsimring:1996p19688,Rouzine:2003p33590,Desai:2007p954,Neher:2010p30641}. The study of fluctuations in the approximately stationary state of mutation selection balance which we present here is a step towards more general quantitative theory of fitness fluctuations in adapting populations. 

\section{Model and Methods}
We assume that mutations happen at rate $\mut$ and that each mutation reduces the growth rate of the genotype by $s\ll 1$. Within this model, proposed and formalized by \citeauthor{Haigh:1978p37141}, individuals can be categorized by the number of deleterious mutations they carry. The equation describing the fitness distribution in the population, i.e., what part $\n_k$ of the population carries $k$ deleterious mutations, is given by
\begin{equation}
\label{eq:model}
\frac{d}{d t}  \n_k= s(\bar{k}-k)\n_k-\mut\n_k + \mut\n_{k-1} + \sqrt{\n_k}\eta_k
\end{equation}
where $\bar{k} = N^{-1}\sum_k k \n_k$ ($\sum_k \n_k=N$) and the last term accounts for fluctuations due to finite populations, i.e., genetic drift, and has the properties of uncorrelated Gaussian white noise with $\la \eta_k(t)\eta_l(t')\ra = \delta_{kl}\delta(t-t')$. In the infinite population limit, this equation has the well known steady state solution $\ns_k = Ne^{-\lambda} \lambda^k/k!$, where $\lambda = \mut/s$. A time dependent analytic solution of the deterministic model has been described in \citep{Etheridge:2007p44291}.

Note that we have deviated slightly from the standard model, which assumes that genetic drift amounts to a binomial resampling of the distribution with the current frequencies $N^{-1}\n_k$. This choice would result in off-diagonal correlations between noise terms that stem from the constraint that the total population size is strictly constant. This exact population size constraint is an arbitrary model choice which we have relaxed to simplify the algebra. Instead, we control the population size by a soft constraint which keeps the population constant on average but allows small fluctuations of $N$. The implementation of this constraint is described explicitly below. We confirmed the equivalence of the two models by simulation.

\subsection*{Computer simulations}
We implemented the model as a computer simulation with discrete generations, where each generation is produced by a weighted resampling of the previous generation. Specifically, 
\begin{equation}
\n_k(t+1) \leftarrow \mathrm{Poisson}\left(\frac{1}{\bar{W}}\sum_{i=0}^k \frac{e^{-\mut} \mut^i}{i!}(1-s)^{k-i}\n_{k-i}(t)\right)
\end{equation}
where $\bar{W}$ is the mean fitness $\bar{W}=C N^{-1}\sum_k (1-s)^k \n_k$ and $C = \exp(NN_0^{-1}-1)$ is an adjustment made to the overall growth rate to keep the population size approximately at $N_0$. 
This specific discretization is chosen because it has exactly the same stationary solution as the continuous time version above \citep{Haigh:1978p37141}.  The simulation was implemented in Python using the scientific computing environment SciPy \citep{Oliphant:2007p25672}. If the parameter of the Poisson distribution was larger than $10^4$, a Gaussian approximation to the Poisson distribution was used to avoid integer overflows. 

To determine the ratchet rate, the population was initialized with its steady state expectation $\ns_k$, allowed to equilibrate for $10^4$ generations, and then run for further $T=10^8$ generations. Over these $10^8$ generations, the number of clicks of the ratchet where recorded and the rate estimated as clicks per generation. 

The source code of the programs, along with a short documentation, is available as supplementary material. In addition, we also provide some of the raw data and the analysis scripts producing the figures as they appear in the manuscript.

\subsection*{Numerical determination of the most likely path.}
The central quantities in our path integral formulation of the rate of Muller's ratchet are i) the most likely path to extinction and ii) the associated minimal action $\Smin_{\lambda}$. 
To determine the most likely path to extinction of the fittest class, we discretize the trajectory into $m$ equidistant time points $\rho_i$ between $0$ and $\tau$, where $\x_0(0)=\xs_0$ and $\x_0(\rho_m)=0$.  For a given set of $\x_0(\rho_i)$, a continuous path $\x_0(\rho)$ is generated by linear interpolation. For a given trajectory $\x_0(\rho)$, we determine the mean fitness by solving the deterministic equations for $\x_k(\rho)$, $k\geq 1$. Note that in this scheme, the only independent variable is the path $\x_0(\rho)$, all other degrees of freedom are slaved to $\x_0(\rho)$. From $\x_0(\rho)$ and the resulting $\mk(\rho)$, we calculate the action $S_{\lambda}(\{\x_0(\rho)\})$ as defined in \EQ{path_integral}. $S_{\lambda}(\{\x_0(\rho)\})$ is then minimized by changing the values of $\x_0(\rho_i)$, $0<i<m$, using the simplex minimization algorithm implemented in SciPy \citep{Oliphant:2007p25672}. To speed up convergence, the minimization is first done with a small number of pivot points ($m=4$), which is increased in steps of 2 to $m=24$. The total time $\tau=20$ (in units of $s^{-1}$) was used which is sufficiently large to make the result independent of $\tau$. The code used for the minimization is provided as supplementary material.

\begin{table}
\begin{tabular}{|l|l|}
\hline
Symbol & Description \\
\hline \hline
$N$, $u$, $s$ & population size, mutation rate, and mutation effect \\
$\n_k(t)$, $\ns_k$ & number of individuals in class $k$ at time $t$, steady state value\\
$\lambda = u/s$, $\tau=ts$ & rescaled mutation rate and time \\
$\x_k(\tau)$, $\xs_k$ &  population frequency in class $k$ at time $\tau$, steady state value \\ 
$\xz_{\tau}$, $\xzs$ & abbreviation for $\x_0(\tau)$ and $\xs_0$ \\
$\mk$ & mean fitness: $\mk=\sum_k k\x_k$\\
$\dx_k$, $\dk$&  deviations from steady state: $\dx_k = \x_k-\xs_k$, $\dk=\lambda -\mk$\\
$S_{\lambda}(\{\xz_\rho\})$ & path integral action depending on the path $\{\xz_{\rho}\}$ with $0\leq \rho \leq\tau$.\\
$\Smin_{\lambda}(\xz_{\tau}, \xz_{0})$, $\xz^*_{\rho}$ & extremal action and the associated path depending on the endpoints $\xz_{\tau}$,$\xz_{0}$\\
$\Smin_{\lambda}(\xz)$ & long time limit of $\Smin_{\lambda}(\xz_{\tau}, \xz_{0})$ with $\xz=\xz_{\tau}$\\
$P_{\tau}(\xz_\tau|\xz_0)$ & propagator from $\xz_0$ to $\xz_\tau$ in time $\tau$\\
$p(\xz)$ & steady state distribution of $\xz$\\
$\rate$ & rate of the ratchet in units of $s$ \\
$\sigma^2$ & variance of $\x_0$ depending on $Ns$ and $\lambda$\\
$\zeta^2$ & rescaled variance of $\x_0$ depending on $\lambda$ only: $\zeta^2 = Nse^{\lambda}\sigma^2$\\
$\mr{i}_k$, $\ml{i}_k$ & $k$-th component of right and left eigenvectors with eigenvalue $-i$\\
$a_i(\tau)$ & projection of $\dx_k$ on $\ml{i}_k$\\
$\alpha$ & parameter of the effective potential confining $\x_0$ (traditionally  $\alpha=0.5$-$0.6$) \\
\hline
\end{tabular}
\caption{List of symbols}
\label{tablelabel}
\end{table}

\section{Results and Discussion}
Fluctuations of the size $\n_0$ of the least loaded class can lead to its extinction. In the absence of beneficial mutations this class is lost forever \citep{Muller:1964p45018}, and the resulting accumulation of deleterious mutations could have dramatic evolutionary consequences. Considerable effort has been devoted in understanding this process, and it has been noted that the rate at which the fittest class is lost depends strongly on the average number of individuals in the top class $\ns_0$ \citep{Haigh:1978p37141}. Later studies have shown that the rate is exponentially small in $\ns_0 s$ if $\ns_0 s \gg 1$ \citep{Jain:2008p45047}. If $\ns_0 s$ is small, the ratchet clicks frequently and a traveling wave approach is more appropriate \citep{Rouzine:2008p20864}. However, a quantitative understanding of the $\ns_0s\gg 1$ regime is still lacking.

Here, we present a systematic analysis of the problem by first analyzing how selection stabilizes the population against the destabilizing influences of mutation and genetic drift, and later use this insight to derive an approximation to the rate of Muller's ratchet. Before analyzing \EQ{model}, it is useful to realize that it implies a common unit of time for the time derivative, the mutation rate and the selection coefficient which is of our choosing (days, months, generations, \ldots). We can use this freedom to simplify the equation and reveal what the important parameters are that govern the behavior of the equation. In this case, it is useful to use $s^{-1}$ as the unit of time and work with the rescaled time $\tau = ts$. Furthermore, we will formulate the problem in terms of frequencies $\x_k = N^{-1}\n_k$ rather than numbers of individuals, and obtain
\begin{equation}
\label{eq:model_rescaled}
\frac{d}{d \tau}  \x_k= (\bar{k}-k-\lambda)\x_k + \lambda \x_{k-1} + \sqrt{\frac{\x_k}{Ns}}\eta_k
\end{equation}
where $\lambda = \mut/s$ is the dimensionless ratio of mutation rate and selection strength. In other words, $\lambda$ is the average number of mutations that happen over a time $s^{-1}$ (our unit of time). Note that $\lambda$ uniquely specifies the deterministic part of this equation and its steady state solution $\xs_k = \lambda^k e^{-\lambda}/k!$. The stochastic forces are proportional to $1/\sqrt{Ns}$. Again, the parameter combination $(Ns)^{-1}$ has a simple interpretation as the variance of the stochastic effects accumulated over time $s^{-1}$.
Other than through a prefactor determining the unit of time, any quantity governed by \EQ{model_rescaled} can depend only on $\lambda$ and $Ns$. Hence it is immediately obvious that the ratchet rate cannot depend on $\ns_0 = Ne^{-\lambda}$ alone, but has to depend on $\ns_0 s$ instead \citep{Jain:2008p45047}.  All times and rates in rescaled time units are denoted by greek letters, while we  use roman letters for times and rates in units of generations. 

Before turning to the ratchet rate, we shall analyze in greater detail the interplay of deterministic and stochastic forces in \EQ{model_rescaled}. A full time dependent analytic solution of the deterministic model was found in \citep{Etheridge:2007p44291}. Below, we will present an analytic characterization of the stochastic properties of the system in a limit where stochastic perturbations are small.

\begin{figure*}[htb]
\begin{center}
  \includegraphics[width=0.48\columnwidth]{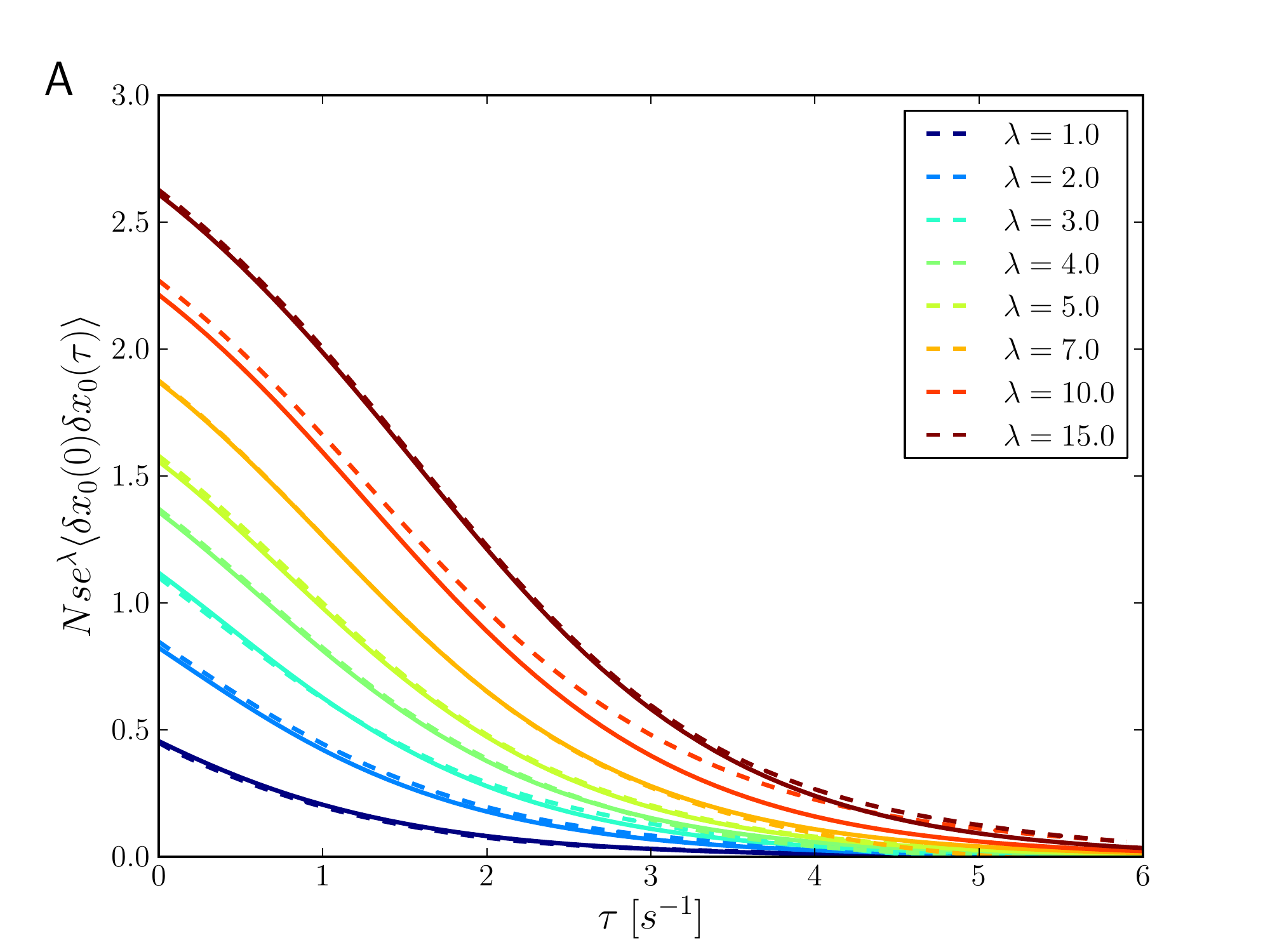}
  \includegraphics[width=0.48\columnwidth]{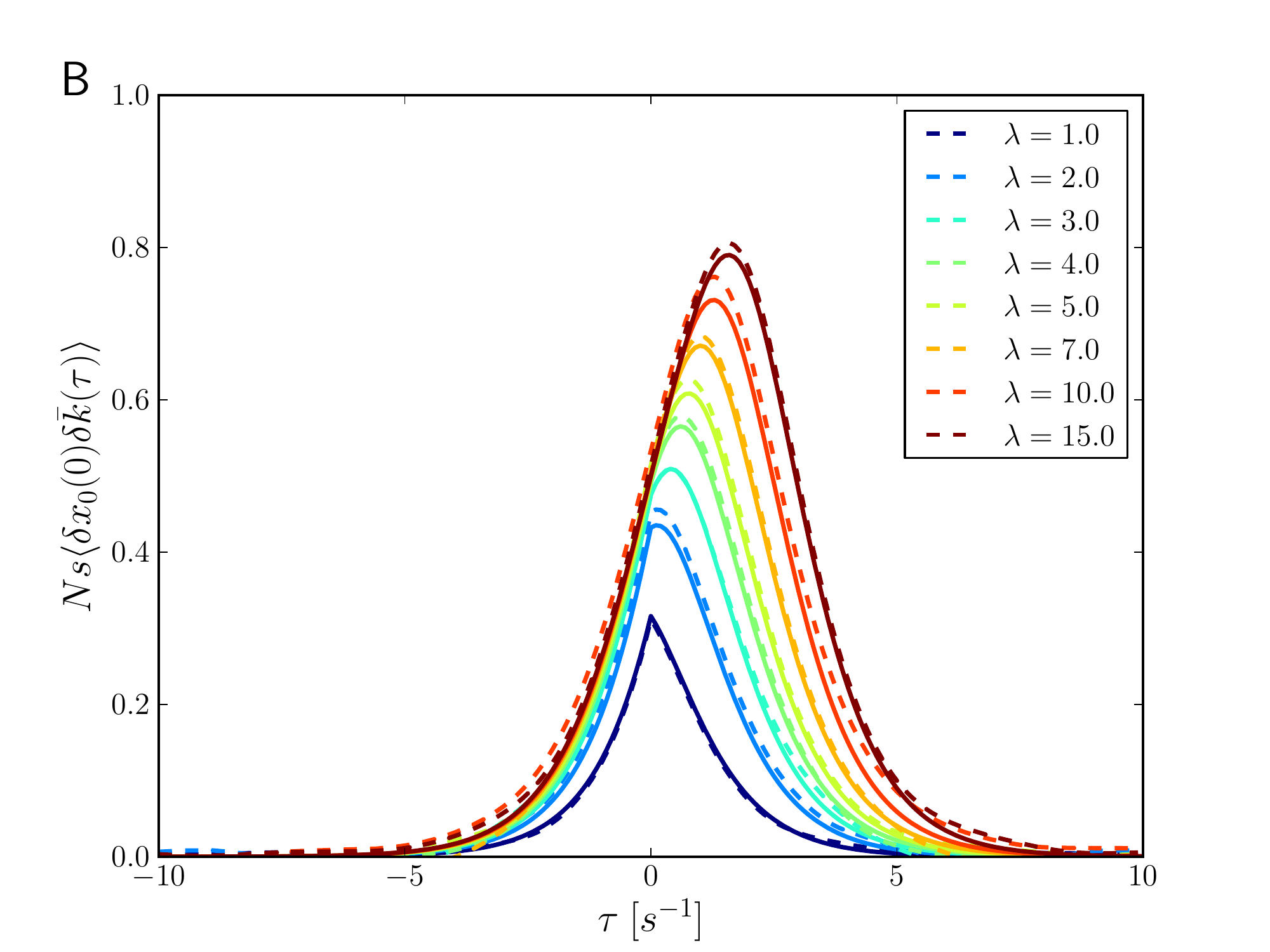}
  \caption[labelInTOC]{Panel A: the covariance of the size of the fittest class $\x_0(0)$ with $\x_0(\tau)$ a time $\tau$ later. The normalized auto-correlation of $\x_0$ increases with $\lambda$. Panel B: the covariance of $\x_0(0)$ with the mean fitness at time $\tau$ in the past or future. One observes a pronounced asymmetry, showing that fluctuations of the fittest class propagate towards the bulk of the fitness distribution and results in delayed fluctuations. Simulation results are shown as dashed lines, theory curves are solid. In all cases, $s=0.01$ and $N\x_0 s=100$. Note that time is measured in units of $1/s$, which is the natural time scale of the dynamics. }
  \label{fig:correlation_functions}
\end{center}
\end{figure*}

\subsection*{Linear stability analysis.} 
In the limit of large populations, the fluctuations of $\x_k$ around the deterministic steady state $\xs_k$ can be analyzed in linear perturbation theory. In other words, we express deviations from the steady state as $\dx_k =\x_k -\xs_k$ and expand the deterministic part of \EQ{model_rescaled} to  order $\dx_k^2$. This expansion
\begin{equation}
\label{eq:linear}
\frac{d}{d\tau} \dx_k= -k \dx_k +\lambda \dx_{k-1}+\xs_k\sum_{m=0} (m-\lambda)\dx_m = \sum_m L_{km}\dx_m
\end{equation}
defines a linear operator $L_{km}$. A quick calculation shows that $L_{km}$ has eigenvalues $\kappa_i = -i$ with $i=0, 1, 2\ldots$. The right eigenvector corresponding to $\kappa_0=0$ is simply $\mr{0}_k = \xs_k$, while the right eigenvectors for $i>0$ are given by 
\begin{equation}
\label{eq:eigenmodes}
\mr{i}_k=\xs_{k-i} - \xs_{k}
\end{equation}
where $k$ numbers the coordinate of the vector. This is readily verified by direct substitution (Note that $\xs_{i}=0$ for $i<0$).

The eigenvector $\mr{0}_k$ corresponds to population size fluctuations which in our implementation are a controlled by a carrying capacity. The eigenvalue associated with this mode in the computer simulation is large and negative and need not be considered here, see Methods.  All other eigenvalues are negative, which is to say that $\xs_k$ is a stable solution.

The eigenvectors for $i>0$ have an intuitive interpretation: Eigenvector $i$ corresponds to a shift of a fraction of the population by $i$ fitness classes downward. Since such a shift reduces mean fitness, the fittest classes start growing, and undo the shift. More generally, any small perturbation of the population distribution can be expanded into eigenvectors $\dx_k(\tau) = \sum_j \mr{j}_k a_j(\tau)$ and the associated amplitudes $a_j(\tau)$ will decay exponentially in time with rate $j$ (remember that the unit of time is $s^{-1}$). Since the amplitudes are projections of $\dx_k$ onto the left eigenvectors of $L_{mk}$, we need to know those as well. For $\kappa_0=0$, the left eigenvector is simply $\ml{0}_k=1$, while the other left eigenvectors are given by
\begin{equation}
\ml{i}_k=\frac{(-1)^{k-i}e^{\lambda}\lambda^{i-k}}{(i-k)!} \quad\quad 0\leq k \leq i \ ,
\end{equation}
and $\ml{i}_k=0$ for $k>i$. With the left and right eigenvectors and the eigenvalue spectrum of the deterministic system on hand, we will now re-instantiate the stochastic part of the dynamics. 
\begin{equation}
\frac{d}{d\tau} \dx_k = \sum_m L_{km}\dx_m + \sqrt{\frac{\xs_k}{Ns}}\eta_k
\end{equation}
Note that we approximated the strength of noise by its value at equilibrium. This approximation is justified as long as we consider only small deviations from the equilibrium. The full $\x_k$ dependent noise term will be reintroduced later when we turn to Muller's ratchet. Substituting the representation of $\dx_k(\tau) = \sum_i \mr{i}_k a_i(\tau)$ and projecting onto the left eigenvector $\ml{j}_k$, we obtain the stochastic equations for the amplitudes
\begin{equation}
\label{eq:modes}
\frac{d}{d\tau} a_j(\tau) = -j a_j(\tau) + \sum_k \ml{j}_k \sqrt{\frac{\xs_k}{Ns}} \eta_k(\tau) \ .
\end{equation}
Each noise term $\eta_k$ contributes to every $a_j$ and induces correlations between the $a_j$, but each amplitude can be integrated explicitely
\begin{equation}
a_j(\tau) = \int_{-\infty}^{\tau}d\tau' e^{-j(\tau-\tau')} \sum_k \ml{j}_k \sqrt{\frac{\xs_k}{Ns}} \eta_k(\tau') \ .
\end{equation}
The covariances of different amplitudes are evaluated in the Supplementary Information and found to be
\begin{equation}
\label{eq:mode_cov}
\begin{split}
\langle a_i(\tau) a_j(\tau+\Delta \tau)\rangle = \frac{e^{-j \Delta \tau }}{i+j}\sum_{k} \frac{\ml{i}_k\ml{j}_k\xs_k}{Ns}
\end{split}
\end{equation}
However, we are not primarily interested in the covariance properties of the amplitudes of eigenvectors, but expect that the fluctuations of the fittest class and fluctuations of the mean fitness are important for the rate of Muller's ratchet and other properties of the dynamics of the population.
To this end we express $\dx_0(\tau)$ and $\bar{k}(\tau)$ as 
\begin{eqnarray}
\dx_0(\tau) &=& \sum_{j>0} \mr{j}_0 a_j(\tau)  = - e^{-\lambda} \sum_{j>0} a_j(\tau)\\
\dk(\tau) &=& -\sum_{j>0,k} k\mr{j}_k a_j(\tau) =  -\sum_{j>0} j a_j(\tau) \ .
\end{eqnarray}
Together with \EQ{mode_cov}, we can now calculate the desired quantities. The calculations required to break down the multiple sums to interpretable expressions are lengthy, but straight-forward and detailed in the supplement. Below, we will present and discuss the results obtained in the supplement.

\subsubsection*{Fluctuations of $\x_0$ and the mean fitness.}
For the variance of the fittest class $\la \dx_0^2\ra$ and more generally its auto-correlation, we find
\begin{equation}
\begin{split}
\label{eq:variance}
&\langle \dx_0(0) \dx_0(\tau)\rangle = 
 \frac{e^{-\lambda}}{Ns} \int_0^1 \frac{d\theta}{\theta} G_\lambda(\theta,\tau)  \\
&G_\lambda(\theta,\tau) = e^{\lambda \theta^2e^{-\tau} -\lambda(1+e^{-\tau}) \theta}-e^{-\lambda \theta}-e^{-\lambda \theta e^{-\tau}}+1
\end{split}
\end{equation}
The variance of the fittest class ($\tau=0$ in the above expression) is therefore $\sigma^2 = \frac{\bar{x}_0}{Ns}\zeta^2(\lambda)$ where $\zeta^2(\lambda) = \int_0^1 \frac{d\theta}{\theta} G_\lambda(\theta,0)$ is the standardized  variance of the top bin, which depends only on $\lambda$. 
For small $\lambda$, it simplifies to $\zeta^2(\lambda)\approx \frac{ 1}{2}\lambda+\mathcal{O}(\lambda^2)$. This limit corresponds to $\xs_0$ close to $1$ with only a small fraction of the population carrying deleterious mutations $\x_1\approx \lambda = \mut/s$. The opposite limit of large $\lambda$ corresponds to a broad fitness distribution where the top class represents only a very small fraction of the entire population. In this limit, the leading behavior of the variance $\sigma^2$ is $\sim \frac{\xs_0\log \lambda}{Ns}$.
The full auto-correlation function is shown in \FIG{correlation_functions}A for different values of $\lambda$ and compared to simulation results, which agree within measurement error. In our rescaled units, the correlation functions decay over a time of order $1$, corresponding to a time of order $1/s$ in real time. More precisely, the decay time (in scaled units) increases with increasing $\lambda$ as $\log \lambda$. 

In a similar manner, we can calculate the auto-correlation of the mean fitness 
\begin{equation}
\begin{split}
&\langle \delta  \bar{k}(0) \delta\bar{k}(\tau)\rangle =\frac{\lambda e^{\lambda}}{N s}\int_0^1 d\theta I_{\lambda}(\theta, \tau)\\
&I_{\lambda}(\theta, \tau) = e^{-\tau} e^{\lambda \theta^2e^{-\tau} -\lambda(1+e^{-\tau}) \theta} \left(\theta  +\lambda \theta\left(\theta e^{-\tau}- 1\right)\left(\theta- 1\right)\right)
 \end{split}
\end{equation}
which asymptotes to $(4Ns \xs_0)^{-1}$ for large $\lambda$ at $\tau=0$. It is hence inversely proportional to the size of the fittest class $\x_0$. For large $\lambda$, $\x_0$ represents only a tiny fraction of the population and fluctuations of the mean can be substantial even for very large $N$. This emphasizes the importance of fluctuations of the size of the fittest class for properties of the distribution. 

If fluctuations of the mean fitness $\bar{k}$ are driven by fluctuations of the fittest class $\x_0$, we expect a strong correlation between those fluctuations \citep{Etheridge:2007p44291}. Furthermore, fluctuations of $\x_0$ should precede fluctuations of the mean. These expectations are confirmed by the analytic result 
\begin{equation} 
\langle \dx_0(0) \dk(\tau)\rangle = \frac{\lambda}{Ns} \int_0^1 d\theta H_\lambda(\theta,\tau)
\end{equation} 
where
\begin{equation}
H_\lambda(\theta,\tau) = 
\begin{cases} 
(\theta-1)e^{-\tau+e^{-\tau}\lambda \theta^2-\lambda(1+e^{-\tau}) \theta}+ e^{-\tau-e^{-\tau}\lambda \theta} & \tau >0 \\
(e^{\tau} \theta-1)e^{e^{\tau}\lambda \theta^2-\lambda(1+e^{\tau}) \theta}+e^{-\lambda \theta} & \tau <0 
\end{cases}
\end{equation}
This expression is shown in \FIG{correlation_functions}B for different values of $\lambda$. The cross correlation $\langle \dx_0(0) \dk(\tau)\rangle$  is asymmetric in time: With larger $\lambda$, the peak of the correlation function moves slowly (logarithmically) to larger delays. This result is intuitive, since we expect that fluctuations in the fittest class will propagate to less and less fit classes and that the dynamics of the entire distribution is, at least partly, slaved to the dynamics of the top class.

In all of these three cases, the magnitude of the fluctuations is governed by the parameter $Ns$, while the shape of the correlation function depends on the parameter $\lambda$. Only the unit in which time is measured has to be compared to the strength of selection directly.

\subsection*{The rate of Muller's Ratchet.}
\begin{figure}[tp]
\begin{center}
  \includegraphics[width=0.99\columnwidth]{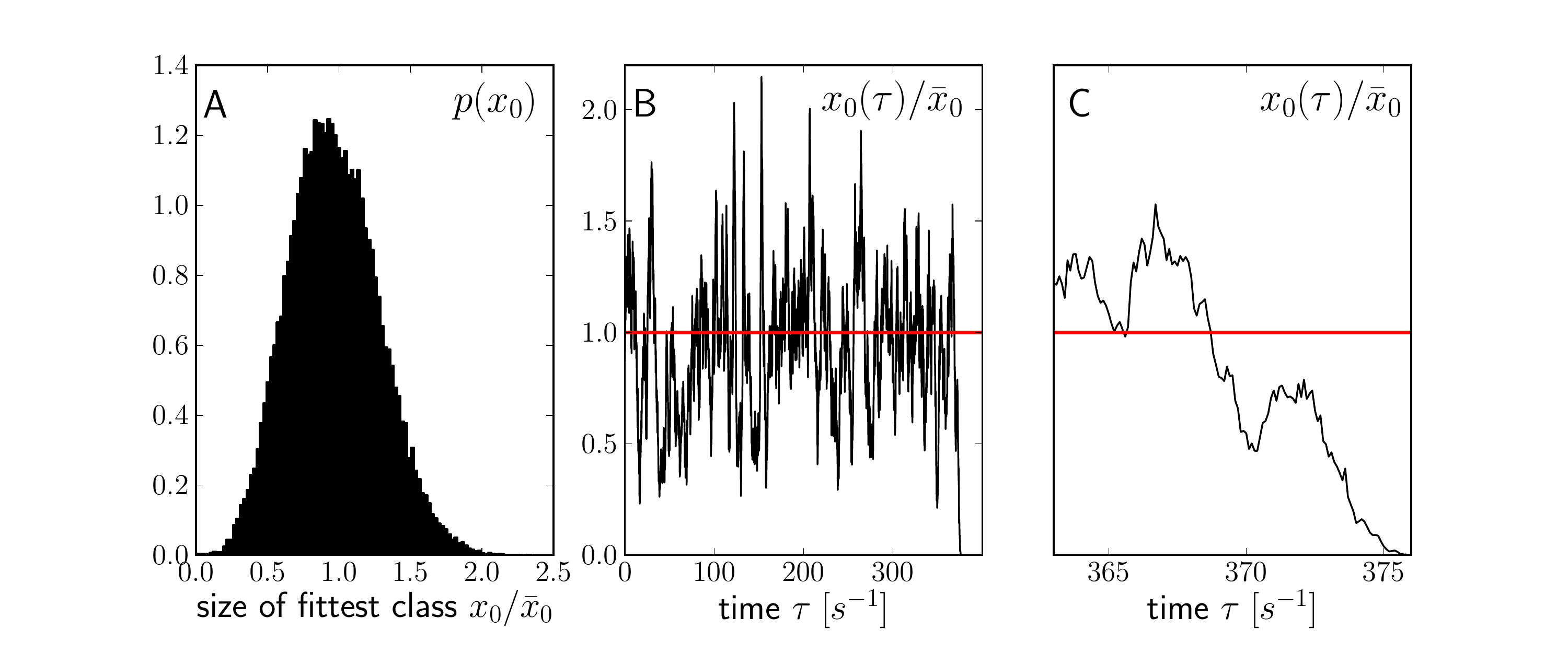}
  \caption[labelInTOC]{An example of a click of the ratchet with $N=5\times 10^{7}$, $s=0.01$ and $\lambda=10$, corresponding to an average size of the fittest class $\ns \approx 2269$. Panel A: The distribution of $\x_0$ averaged over the time prior to extinction. Panel B\&C: The trajectory of $\x_0(\tau)$, with the part of the trajectory that ultimately leads to extinction magnified in panel C. The final run towards $\x_0=0$ takes a few time units, as expected from the results for the correlation functions which suggest a (rescaled) correlation time of $\sim \log \lambda$. Note this time corresponds to a few hundred generations since $s=0.01$.}
  \label{fig:click_illustration}
\end{center}
\end{figure}

The ratchet clicks when the size of the fittest class hits 0, and the rate of the ratchet is given by the inverse of the mean time between successive clicks of the ratchet. Depending on the average size $\xs_0$ of the fittest class, the model displays very different behavior. If $Ns\xs_0$ is comparable to or smaller than 1, the ratchet clicks often without settling to a quasi-equilibrium in between clicks. This limit has been studied in \citep{Rouzine:2008p20864}. Conversely, if $Ns\xs_0 \gg 1$ ratchet clicks are rare and the system stays a long time close to its quasi-equilibrium state $\xs_k$. Such a scenario, taken from simulations, is illustrated in \FIG{click_illustration}. Panel A shows the distribution of $\x_0$ prior to the click, while  \FIG{click_illustration}B shows the realized trajectory which ends at $\x_0=0$. Prior to extinction, $\x_0(\tau)$ fluctuates around its equilibrium value and large excursions are rare and short. The final fluctuation which results in the click of the ratchet is zoomed in on in \FIG{click_illustration}C. Compared to the time the trajectory spends near $\xs_0$, the final large excursion away from the steady is short and happens in a few units of rescaled time. Translated back to generations, the final excursion took a few hundred generations ($s=0.01$ in this example).

In rescaled time, the equation governing the frequency of the top class is
\begin{equation}
\label{eq:topbin_langevin_nonlocal}
\frac{d}{d\tau} \x_0(\tau) = \dk(\tau)\x_0(\tau)+ \sqrt{\frac{\x_0(\tau)}{Ns}}\eta_0(\tau)
\end{equation}
where $\dk(\tau)=\lambda - \mk(\tau)$.
The restoring force $\dk$ depends on $\x_0$, as well as on the size of the other classes $\x_k$. For sufficiently large $\lambda$,  $\x_0$ is much smaller than $\x_k$ with $k\geq 1$, such that the stochastic force is most important for $\x_0$. The dynamics of $\x_k$, $k\geq1$, is approximately slaved to the stochastic trajectory of $\x_0(\tau)$. We can therefore try to find an approximation of $\dk(\tau)$ in terms of $\x_0(\tau)$ only. The linear stability analysis of the mutation selection balance has taught us that the restoring force exerted by the mean fitness on fluctuations in $\x_0$ is delayed with the delay increasing with $\propto \log \lambda$. The latter observation implies that the restoring force on $\x_0$ will mainly depend on the values of $\x_0$ some time of order $\log \lambda$ in the past. Such history dependence complicates the analysis, and this delay has been ignored in previous analysis, which assumed that $\dk(\tau)$ depends on the instantaneous value of $\x_0(\tau)$ via $\dk(\tau)= \alpha (1-\x_0(\tau)/\xs_0)$ \citep{Stephan:1993p42929,Gordo:2000p42688,Jain:2008p45047}. The parameter $\alpha$ was chosen ad hoc between $0.5$ and $0.6$. This restoring force is akin to an harmonic potential centered around $\xs_0$ and the stochastic dynamics is equivalently described by a diffusion equation for the probability distribution $P(\x_0,\tau)$. 
\begin{equation}
\label{eq:diffusion}
\frac{\partial}{\partial \tau}P(\xz,\tau) = \frac{1}{2Ns}\frac{\partial^2}{\partial \xz^2} \xz P(\xz,\tau) - \alpha\frac{\partial}{\partial \xz}  (1-\xz/\xzs)\xz P(\xz,\tau)
\end{equation}
where we have denoted $\x_0$ by $\xz$ for simplicity. The fact that the fittest class is lost whenever its size hits 0 corresponds to an absorbing boundary condition for $P(\xz,\tau)$ at $\xz=0$. For such a one dimensional diffusion problem, the mean first passage time can be computed in closed form \citep{WGardiner:2004p36981} and this formula has been used in \citep{Stephan:1993p42929,Gordo:2000p42688} to estimate the rate of the ratchet. An accurate analytic approximation to that formula has been presented by \citet{Jain:2008p45047}. For completeness, we will present an alternative derivation of these results that will help to interpret the more general results presented below.
In the limit of interest, $Ns\xzs\gg 1$, clicks of the ratchet occur on much longer time scales than the local equilibration of $\xz$. We can therefore approximate the distribution as $P(\xz,\tau)\approx e^{-\rate \tau} p(\xz)$ where $\rate$ is the rate of the ratchet. In this factorization, $p(\xz)$ is the quasi-steady distribution shown in \FIG{click_illustration}A, while $\rate$ is the small rate at which $P(z,\tau)$ looses mass due to events like the one shown in \FIG{click_illustration}C. Inserting this ansatz and integrating \EQ{diffusion} from $\xz$ to $\infty$, we obtain
\begin{equation}
\label{eq:steady_state_diffusion}
-\rate P(X>\xz) = \frac{1}{2Ns} \frac{\partial}{\partial\xz} \xz p(\xz) - \alpha \xz(1-\xz/\xzs) p(\xz)
\end{equation}
 where $P(X>\xz) = \int_\xz^\infty d\xz' p(\xz')$, which is $\approx 1$ for $\xz < \xzs$ and rapidly falls to $0$ for $\xz>\xzs$. To obtain the rate $\rate$, we will solve this equation in a regime of small $\xz \ll \xzs$, where the term on the left is important but constant, and in a regime $\xz \gg (Ns)^{-1}$, where the term on the left can be neglected. For the general discussion below, it will be useful to solve this equation for a general diffusion constant $D(\xz)$ (here equal to $\xz/2Ns$), force field $A(\xz)$ (here equal to $\alpha \xz(\xz/\xzs-1)$), and a constant $C$
\begin{equation}
\label{eq:general}
-C = \frac{\partial}{\partial\xz} D(\xz) p(\xz) + A(\xz) p(\xz)
\end{equation}
with solution 
\begin{equation}
\label{eq:diffusion_general}
p(\xz) = \frac{1}{D(\xz)} e^{-\int_0^\xz dy \frac{A(y)}{D(y)} }\left[\beta - C\int_0^\xz dy e^{\int_0^y dy' \frac{A(y')}{D(y')} }\right]
\end{equation}
Note that this solution is inversely proportional to the diffusion constant, while the dependence on selection is accounted for by the exponential factors.
For $\xz\ll\xzs$, $C=\rate$ and $2\alpha Ns\int_0^\xz dy \;(1-\xz/\xzs) \approx 2\alpha Ns \xz$, such that
\begin{equation}
\label{eq:b_layer}
p(\xz) \approx {\rate} {e^{2Ns \alpha \xz} -1 \over \alpha \xz} \quad\quad \xz\ll \xzs \ ,
\end{equation}
where $\beta$ is fixed by the boundary condition that $p(\xz)$ is finite at $\xz=0$.
Note that $\rate=p(0)/2Ns$ relates the rate of extinction to $p(0)$. To determine the latter we need to match the $\xz\ll\xzs$ regime to the bulk of the distribution $\xz \approx \xzs$. As can be seen from \EQ{b_layer}, the constant term $C$ is unimportant in this regime ($e^{2Ns\alpha \xz}\gg 1$). Setting $C=0$ in \EQ{diffusion_general}, we find
\begin{equation}
\label{eq:approximate_steady_state}
p(\xz) \approx \frac{\xzs}{\xz}\frac{\sqrt{ N s \alpha}}{\sqrt{\xzs \pi }} e^{-\alpha Ns \frac{(\xz-\xzs)^2}{\xzs}} \quad\quad \xz N s \alpha \gg 1 \ .
\end{equation}
The integration constant $\beta$ in \EQ{diffusion_general} is fixed by the normalization. Since $p(\xz)$ is concentrated around $\xz=\xzs$ and has a Gaussian shape around $\xzs$, the normalization factor is simply $1/\sqrt{2\pi \sigma^2}$, where $\sigma^2 = \xzs/2\alpha Ns$ is the variance of the Gaussian. The factor $\xzs/\xz$ corresponds to the $1/D(\xz)$ term, scaled so that it equals 1 in the vicinity of $\xzs$. Note that we have already calculated the variance of $p(\xz)$ earlier, \EQ{variance}, and that consistency with this result would require that $\alpha$ is determined by \EQ{variance}.

The two approximate solutions \EQ{b_layer} and \EQ{approximate_steady_state} are both accurate in the intermediate regime $(N s \alpha)^{-1}\ll\xz \ll \xzs$, which allows us to determine the rate $\rate$ in \EQ{b_layer} by matching the two solutions. This matching implies that 
\begin{equation}
\label{eq:simple_ratchet}
\rate = \frac{\sqrt{\xzs Ns \alpha^3}}{\sqrt{\pi}} e^{-\alpha Nse^{-\lambda}}
\end{equation}
which agrees with results obtained previously \citep{Jain:2008p45047}. Note that this rate only depends on the parameters $\lambda$ and $Ns$ of the rescaled model. Since rates have units of inverse time, this expression has to be multiplied by $s$ to obtain the rate in units of inverse generations. 

\begin{figure}[tb]
\begin{center}
  \includegraphics[width=0.48\columnwidth]{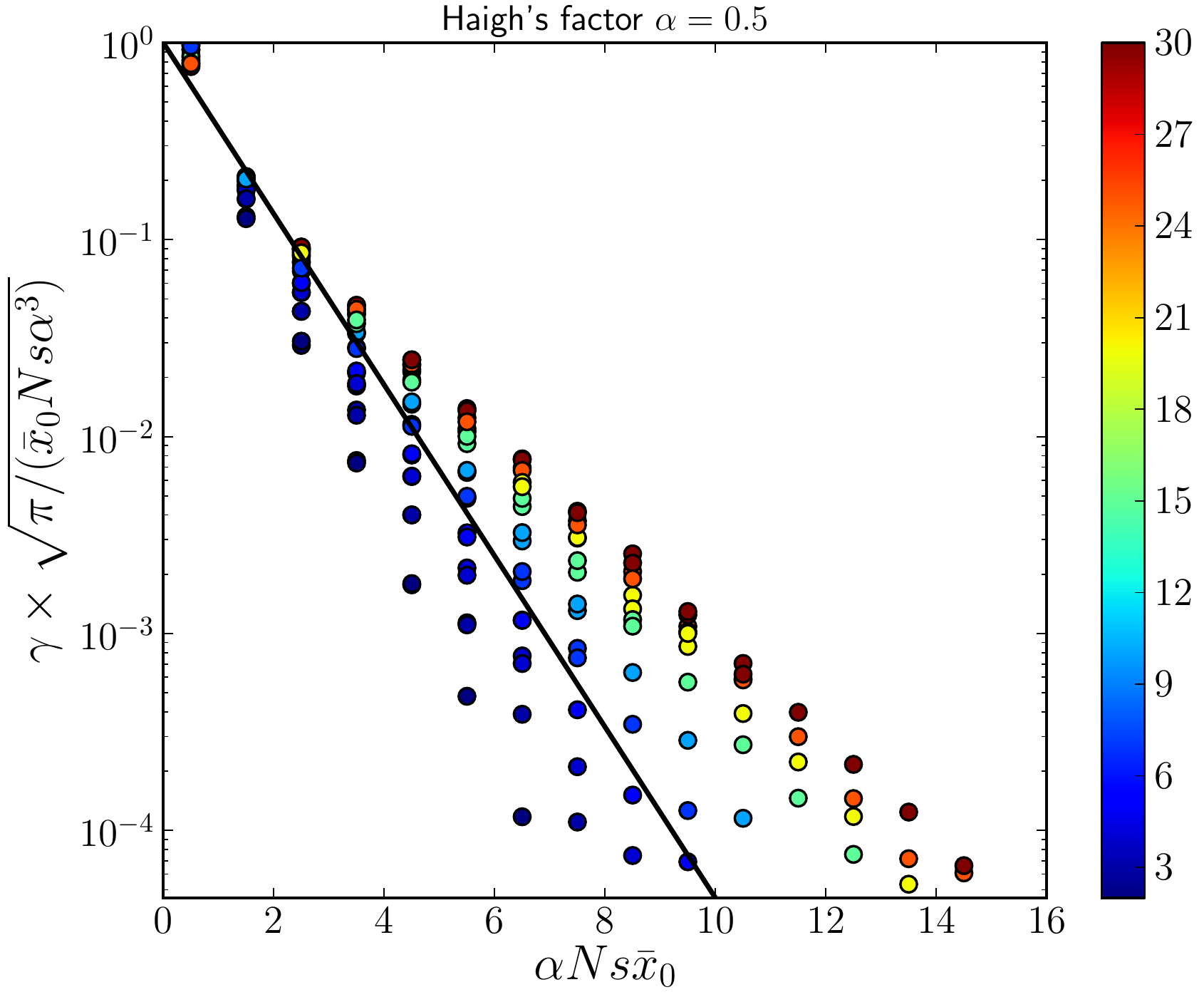}
  \includegraphics[width=0.48\columnwidth]{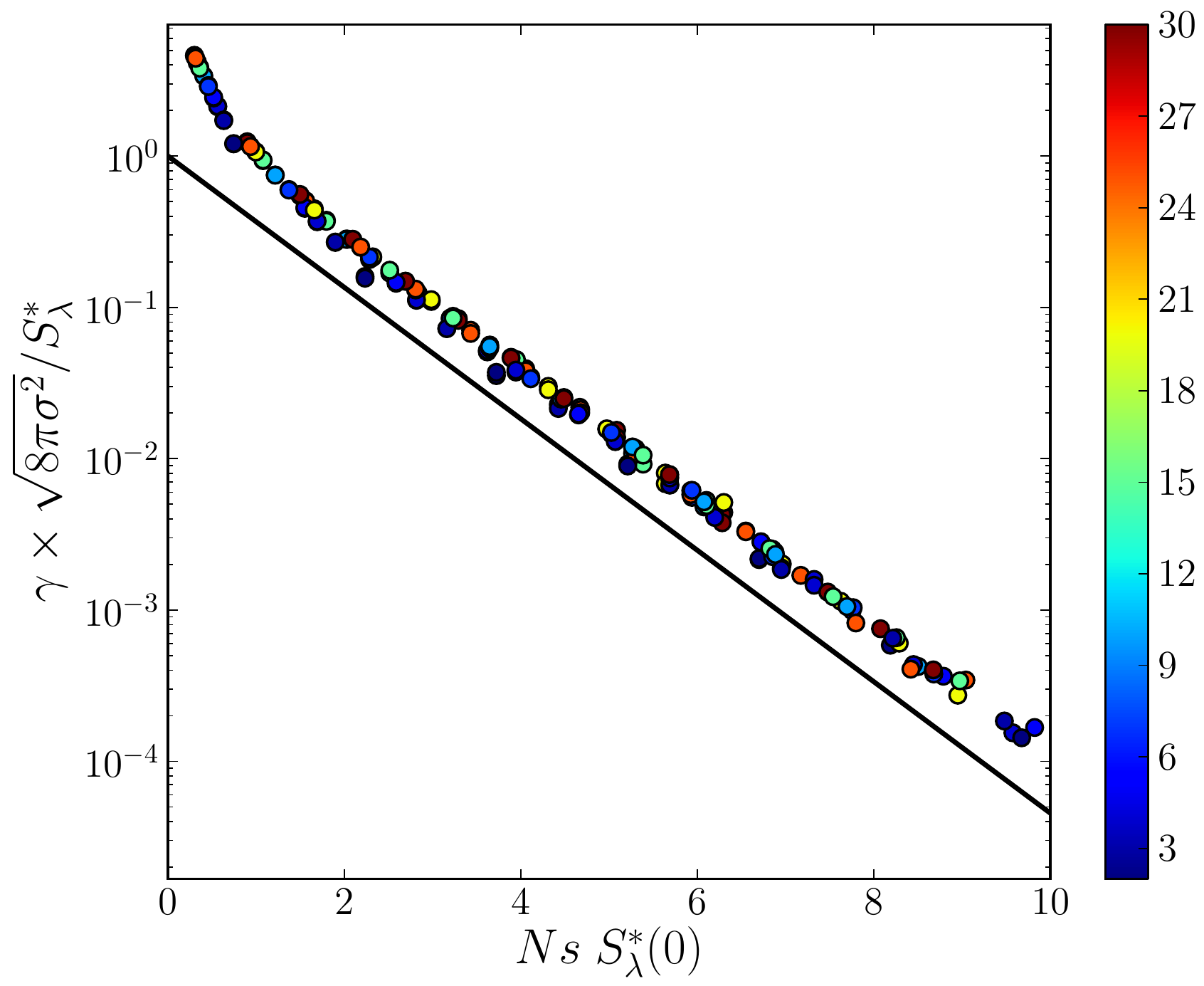}
  \caption[labelInTOC]{The ratchet rate from simulation vs prediction. Both panels show the ratchet rate $\rate$, rescaled with a prefactor to isolate the exponential dependence predicted by analytic approximations; $\lambda$ is color-coded. Panel A compares the simulation results to the prediction of \EQ{simple_ratchet}, which is shown as a straight line. The approximation works only for a particular value of $\lambda$, for otherwise the exponential dependence on $N\xs_0 s$ is not predicted correctly. Panel B compares simulation results to the prediction of \EQ{rate}, again indicated by the straight line. The exponential dependence of rate on $Ns\Smin_{\lambda}(0)$ is well confirmed by simulation results.}
  \label{fig:action}
\end{center}
\end{figure}

However, \EQ{simple_ratchet} does not describe the rate accurately, as is obvious from the comparison with simulation results shown in \FIG{action}A. The plot shows the rescaled ratchet rate $\rate \times\sqrt{\pi}/\sqrt{\xzs N s\alpha^3}$, which according to \EQ{simple_ratchet} should be simply $\exp[-\alpha Nse^{-\lambda}]$, indicated by the black line.  The plot shows clearly that the simulation results often differ from the prediction of \EQ{simple_ratchet} by a large factor. It seems as if $\alpha$ needs to depend on $\lambda$, as we already noticed above when comparing the variance of $p(\xz)$ to \EQ{variance}. In fact, fixing $\alpha$ via \EQ{variance} improves the agreement substantially, but still does not describe the simulations quantitatively.

The reason for the discrepancy is the time delay between $\dk$ and $\xz$ which we quantified by calculating the correlation between $\dk(\tau)$ and $\xz(\tau+\Delta \tau)$. Hence we cannot use an approximation where $\dk$ depends on the instantaneous value of $\xz$, but have to calculate $\dk$ from the past trajectory of $\xz$. If the fittest class is the only one that is strongly stochastic, we can calulate $\dk(\tau)$ for a given trajectory $\xz(\rho)$, $\rho\leq \tau$ by integrating the deterministic evolution equations for $\x_k$ with $k\geq 1$ with $\xz(\rho)$ as an external forcing. 

\EQ{topbin_langevin_nonlocal} now not only depends on $\xz(\tau)$, but on all $\xz(\rho)$ with $\rho\leq \tau$ and cannot be mapped to a diffusion equation. Nevertheless, it corresponds to a well defined stochastic integral, known as a path integral in physics \cite{Feynman:1965}, which is amenable to systematic numerical approximation. To introduce path integrals, it is useful to discretize \EQ{topbin_langevin_nonlocal} in time and express $\xz(\rho_i)$ in terms of the state at time $\rho_{i-1}=\rho_i-\Delta \tau$ and the earlier time points. For simplicity, we will use the notation $\tx{i}$ for $\xz(\rho_i)$.
\begin{equation}
\label{eq:discrete_langevin}
\tx{i}-\tx{i-1} = \Delta \tau \dk_{i-1} \tx{i-1}+ \sqrt{\frac{\tx{i-1}\Delta \tau}{Ns}}\; \eta_{i-1}
\end{equation}
where $\dk_{i-1}$ depends on all previous time points $\rho_j$ with $j<i$.
In the limit $\Delta \tau\to 0$, this difference equation converges against \EQ{topbin_langevin_nonlocal} interpreted in the It\^o sense since the $\xz$ dependent prefactor of the noise term is evaluated at $\rho_{i-1}$ rather than at an intermediate time point between $\rho_{i-1}$ and $\rho_i$. 
We can express this transition probability  $P_{\tau}(\xz|\tx{0})$ between the initial state $\tx{0}$ and the final state $\xz=\tx{m}$ as a series of integrals over all intermediate states $\tx{i}$ for $0<1<m$. 
\begin{equation}
P_\tau(\xz |\tx{0}) = \int \prod_{i=1}^{m-1} d\tx{i} P_{\Delta \tau}(\xz| \{\tx{j}\}_{j<m}) P_{\Delta \tau}(\tx{m-1}| \{\tx{j}\}_{j<m-1})\cdots  P_{\Delta \tau}(\tx{1}| \tx{0})
\end{equation}
Each of these infinitesimal transitions correspond to solutions of \EQ{discrete_langevin} with $\eta_{i}$ drawn from a standard Gaussian \citep{Lau:2007p45316}. 
Hence 
\begin{equation}
P_{\Delta \tau}(\tx{i}| \{\tx{j}\}_{j<i}) = \frac{\sqrt{Ns}}{\sqrt{2\pi \Delta t \tx{i-1}}} \exp\left[-Ns\frac{\left(\tx{i}-\tx{i-1} - \Delta \tau \tx{i-1}\dk_{i-1}\right)^2}{2\Delta \tau \tx{i-1}} \right]
\end{equation}
In the limit of many intermediate steps and small $\Delta \tau$, the transition probability can therefore be written as 
\begin{equation}
\label{eq:path_integral}
P_\tau(\xz|\tx{0}) = \int \mathcal{D} \tx{\rho} \; \exp\left[-Ns \int_0^\tau d\rho \frac{[\dot{\xz}_{\rho} - \tx{\rho}\dk_\rho]^2}{2\tx{\rho}}\right]  = \int \mathcal{D} \tx{\rho} \; e^{-Ns S_{\lambda}(\{ \tx{\rho} \})}
\end{equation}
where $\mathcal{D} \tx{\rho}$ is the limit of $\prod_{i=1}^m d\tx{i}(2\pi \Delta \tau \tx{i-1}/Ns)^{-1/2}$ known as the path integral measure, and we have replaced the discrete time index by its continuous analog. The path integral extends over all continuous path connecting the endpoints $\tx{0}$ and $\tx{\tau}=\xz$. 
The functional $S_{\lambda}(\{\tx{\rho}\})$ in the exponent closely corresponds to the ``action" in physics \citep{Feynman:1965} which is minimized by classical dynamics. Here minimization of the ``action" defines the most likely trajectory. Note that $S_{\lambda}(\{\tx{\rho}\})$ depends on the entire path $\{ \tx{\rho}\}$ with $0\leq \rho \leq \tau$, while the functional itself only depends on $\lambda$. The strength of genetic drift appears as a prefactor of $S_{\lambda}(\{\tx{\rho}\})$ in the exponent. 

The most likely path $\tx{\rho}^*$ connecting the end-points points $\tx{0}$ and $\tx{\tau}$ in time $\tau$ can be determined either by solving the Euler-Lagrange equations or by numerical minimization, see below. Along with the functional, $\tx{\rho}^*$ depends only on $\lambda$. Given this extremal path, we can parameterize every other path connecting $\tx{0}$ and $\tx{\tau}$ as $\tx{\rho}=\tx{\rho}^* + \delta \tx{\rho}$, where $\delta \tx{\rho}$ vanishes at both endpoints ($\delta \tx{0} = \delta \tx{\tau}=0$). Denoting the minimal action associated with $\tx{\rho}^*$ by  $\Smin_{\lambda}(\tx{\tau} ,\tx{0})$, we have
\begin{equation}
P_{\tau}(\tx{\tau} |\tx{0}) = e^{-Ns\Smin_{\lambda}(\tx{\tau} ,\tx{0})} \int \mathcal{D} \delta \tx{\rho} \; e^{-Ns\; \delta S_{\lambda}(\{\delta\tx{\rho}\}, \tx{\tau}, \tx{0})}  = \mathcal{N}^{-1} e^{-Ns\Smin_{\lambda}(\tx{\tau} ,\tx{0})}
\end{equation}
where $\mathcal{N}^{-1}$ factor is equal to the integral over the fluctuations, which in general depends on $\tx{\rho}^*$. The prefactor $Ns$ in $e^{-Ns\; \delta S(\{\delta\tx{\rho}\}, \tx{\tau}, \tx{0})}$ implies that deviations from the optimal path are suppressed in large populations. If $\delta S(\{\delta \tx{\rho}\}, \tx{\tau}, \tx{0})$ is independent of the final point $\tx{\tau}$,  $\mathcal{N}$ can be determined by the normalizing $P_{\tau}(\tx{\tau} |\tx{0})$ with respect to $\tx{\tau}$. In the general case, calculating the fluctuation integral is difficult, and we will determine it here by analogy to the history independent solution presented above \EQ{diffusion}--(\ref{eq:simple_ratchet}).

If the stochastic dynamics admits an (approximately) stationary distribution,  $P_{\tau}(\tx{\tau} |\tx{0})$ becomes independent of $\tau$ and $\tx{0}$ and coincides with the steady state probability distribution $p(\xz)$. It therefore becomes the analog of \EQ{approximate_steady_state}, which for arbitrary diffusion equations is given by the inverse diffusion constant (the prefactor $\xz^{-1}$), multiplied by an exponential quantifying the trade-off between deterministic and stochastic forces. In this path integral representation, the exponential part is played by $e^{-Ns \Smin_{\lambda}(\xz)}$, where $\Smin_{\lambda}(\xz)$ is a function of the final point $\xz$ and $\lambda$ only. The prefactor is independent of the selection term and can hence be determined through the analogy to the Markovian case discussed above
\begin{equation}
\label{eq:bulk}
p(\xz)\approx \frac{\xzs}{\xz}\frac{1}{\sqrt{2\pi \sigma^2}}e^{-Ns\Smin_{\lambda}(\xz)} \quad\quad Ns \xz \gg 1
\end{equation}
where the normalization is obtained by assuming an approximately Gaussian distribution around the steady state value $\xzs$ and the variance $\sigma^2$ is given by \EQ{variance}. Note that this solution is not valid very close to the absorbing boundary since this boundary is not accounted for by the path integral, at least not without some special care. As in the history independent case discussed above, this approximate distribution should be thought of as the time independent  ``bulk" distribution in $P(\xz,\tau) = e^{-\rate\tau}p(\xz)$. To determine the rate extinction rate $\rate$, we again need to understand how probable it is that a trajectory actually hits $\xz=0$, given that it has come pretty close. 

To this end, we need a local solution of \EQ{topbin_langevin_nonlocal} in the boundary layer $\xz\ll \xzs$ as already obtained for the history independent case in \EQ{b_layer}. Once a trajectory comes close to $\xz=0$, it's fate, i.e., whether it goes extinct or returns to $\xz\approx \xzs$, is decided quickly. Hence we can make an instantaneous approximation for $\dk$ which does depend on the past trajectory, but for the time window under consideration it is simply a constant, $\alpha$, yet to be determined. Having reduced the problem to \EQ{b_layer} we can determine  $\alpha$, and hence $\rate$,  by matching of the boundary solution to \EQ{bulk} in the regime $(Ns)^{-1} \ll \xz \ll \xzs$ where both are accurate. The matching condition is
\begin{equation}
\frac{\xzs}{\xz}\frac{1}{\sqrt{2\pi \sigma^2}}e^{-Ns (\Smin_{\lambda}(0)+\xz\partial_\xz \Smin_{\lambda}(\xz)|_{\xz=0})} = \frac{\rate}{\alpha \xz}e^{2\alpha Ns \xz }
\end{equation}
which determines $\alpha$ and $\rate$ by the matching requirement $2\alpha = -\partial_\xz \Smin_{\lambda}(\xz)|_{\xz=0}$ and
\begin{equation}
\label{eq:rate}
\rate=  \frac{|\partial_\xz \Smin_{\lambda}(\xz)|\xzs e^{-Ns \Smin_{\lambda}(0)}}{2\sqrt{2\pi \sigma^2}} \approx
\frac{\Smin_{\lambda}(0)}{\sqrt{8\pi\sigma^2}}e^{-Ns\Smin_{\lambda}(0)}
\end{equation}
where we approximated $|\partial_\xz \Smin_{\lambda}(\xz)|_{\xz=0} \approx \Smin_{\lambda}(0)/\xzs$. The variance $\sigma^2$ is given by \EQ{variance} and depends on $\lambda$ and $Ns$ as $\sigma^2 = \xzs \zeta^2(\lambda)/Ns$. Note that $\rate$ is in units of $s$ and needs to be multiplied by $s$ for conversion to units of inverse generations.  In contrast to Markovian case above, the variance of the ``bulk" is no longer simply related to strength of selection near the $\xz=0$ ``boundary".

\begin{figure}[tb]
\begin{center}
  \includegraphics[width=0.48\columnwidth]{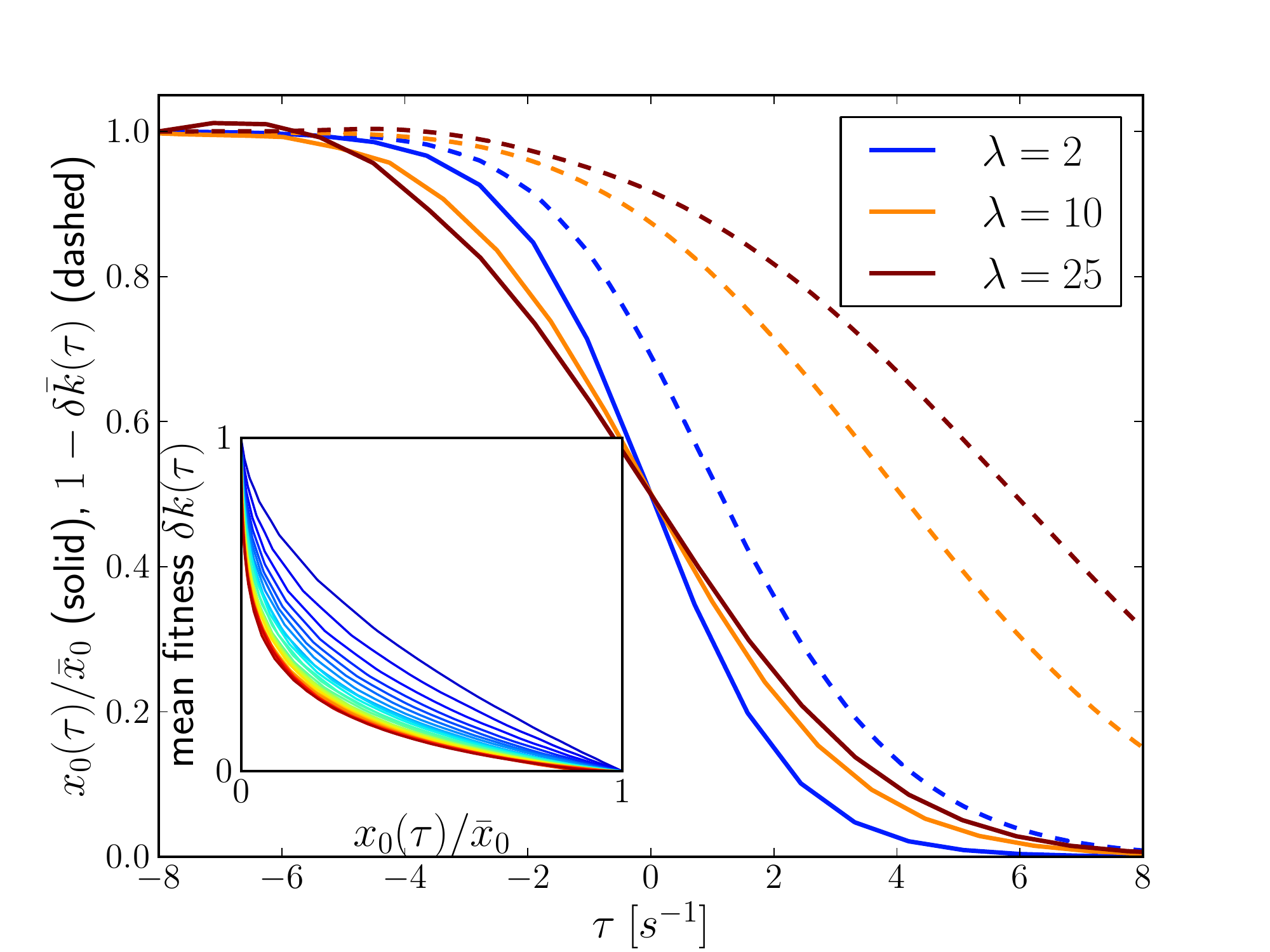}
  \includegraphics[width=0.48\columnwidth]{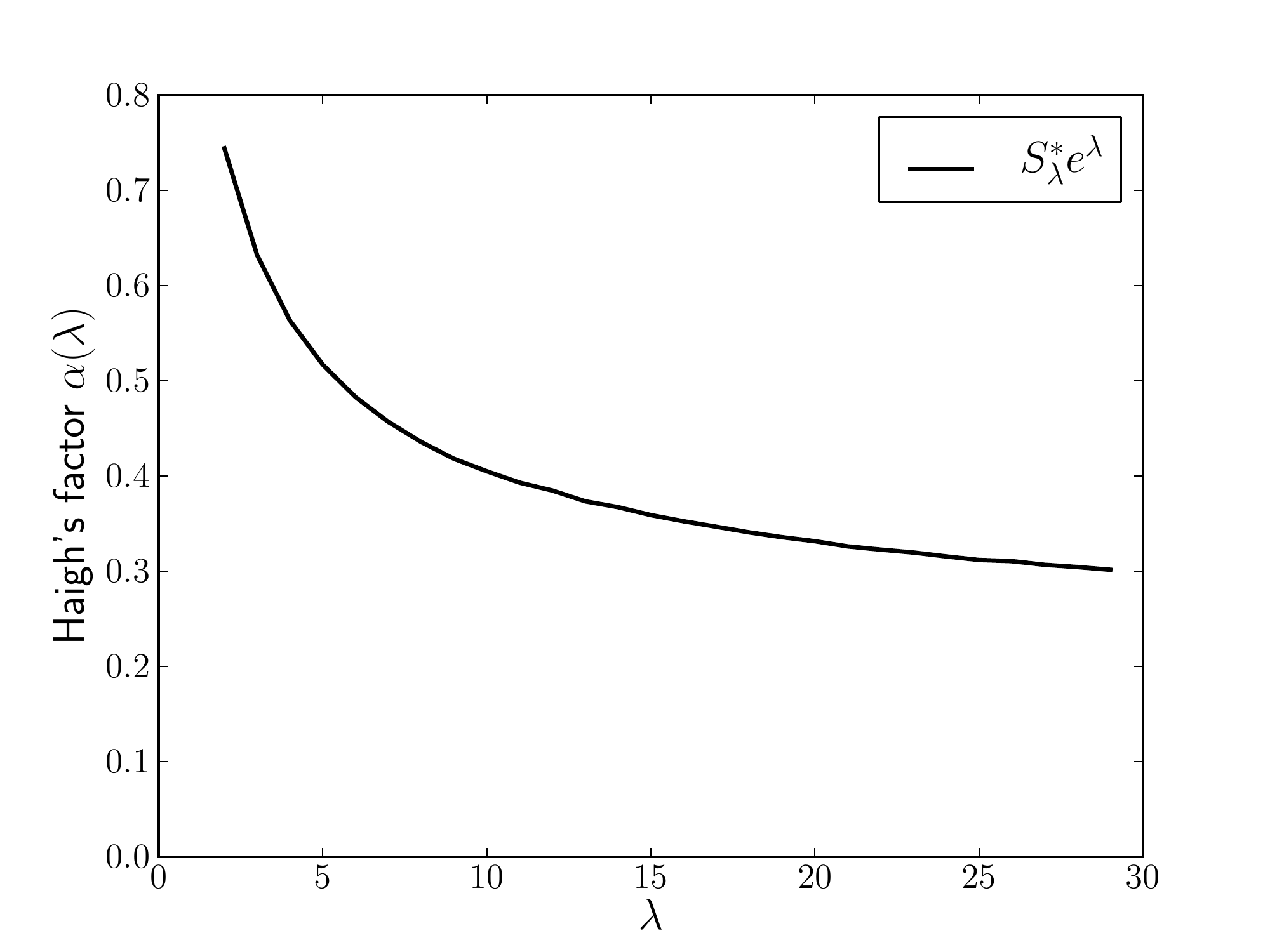}
  \caption[labelInTOC]{Panel A: The most likely path $\x^*_0(\tau)$ to extinction of the fittest class and the concomitant reduction of the mean fitness for different $\lambda$ are plotted against time. Times are shifted such that $\x^*_0(0) = \frac{\xs_0}{2}$. The inset shows the mean fitness $\dk(\tau)$ plotted against $\x^*_0(\tau)$ for different values of $\lambda$. Panel B: ``Haigh's factor'' $\alpha(\lambda) = \Smin_\lambda(0)e^{\lambda}$ as a function of $\lambda$ determined numerically.}
\label{fig:trajectories}
\end{center}
\end{figure}

Since we don't know how to calculate $\Smin_{\lambda}(0)$ or the most likely path $\tx{\rho}^*$ analytically, we determined discrete approximations to $\tx{\rho}^*$ numerically as described in section Model and Methods. Examples of numerically determined most likely path and the corresponding trajectory of the mean fitness are shown in \FIG{trajectories} for different values of $\lambda$. Generically, we find a rapid reduction of the size $\xz$ of the size of the fittest class such that the mean fitness has only partially responded. The inset of \FIG{trajectories}A shows how changes in mean fitness $\dk$ are related to $\xz$ for different $\lambda$. For large $\lambda$, the mean fitness changes only very slowly with $\xz$, which increases the probability of large excursions and hence the rate of the ratchet.

This numerically determined minimal action $\Smin_{\lambda}(0)$ together with the approximation \EQ{rate} describes the rate of the ratchet, as determined in simulations, extremely well. \FIG{action}B shows the same simulation data as \FIG{action}A, but this time rescaled by $\sqrt{8\pi \sigma^2(\lambda)}/\Smin_{\lambda}(0)$ as a function of $Ns\Smin_{\lambda}(0)$. After this rescaling, we expect all data points to lie on the same curve given by $\exp[-Ns\Smin_{\lambda}(0)]$, as is indeed found for many different values of $\mut$, $s$ and $N$ with $\lambda = \mut/s$ ranging from 1 to 30. Note that the vertical shift of the black line relative to the data points depends on the prefactor, which we have approximated. Hence we should not expect agreement better than to about a factor of 2. The important point is that the exponential dependence of the rate on $Ns \Smin_{\lambda}(0)$ is correctly captured by \EQ{rate}.

Previous studies of Muller's ratchet suggested that the rate depends exponentially on $\alpha Nse^{-\lambda}$ \citep{Jain:2008p45047}. To relate this to our results, we determined ``Haigh's factor" $\alpha$ numerically from $\Smin_\lambda(0)$ and plotted it in \FIG{trajectories}B. We find that $\alpha(\lambda)$ drops from around 0.8 to 0.3 as $\lambda$ increases from 1 to 30. The previously used values $0.5-0.6$ for $\alpha$ correspond to $\lambda\approx 6$. Using $\alpha(\lambda) = \Smin_\lambda(0) e^{\lambda}$ as shown in \FIG{trajectories}B, we can recast \EQ{rate} into its traditional form and undo the scaling with $s$. In units of generations, the mean time between clicks is given by
\begin{equation}
T_{click} \approx \frac{2.5\zeta(\lambda)}{\alpha(\lambda)s \sqrt{Nse^{-\lambda}}}e^{Ns \alpha(\lambda)e^{-\lambda}}
\end{equation}
where $\zeta(\lambda)$ is determined by \EQ{variance} and the factor 2.5 is introduced to approximate the part of the prefactor that is independent of $N$, $s$ or $\lambda$. A direct comparison of this expression with simulation results is shown in supplementary figure 1.

\section{Conclusion.}
The main difficulty impeding better understanding of even simple models of evolution is the fact that rare events involving a few or even single individuals determine the fate of the entire population. The important individuals are those in the high fitness tail of the distribution. Fluctuations in the high fitness tail propagate towards more mediocre individuals which dominate a typical population sample. 

We have analyzed the magnitude, decay, and propagation of fluctuations of the fitness distribution in a simple model of the balance between deleterious mutations and selection. In this model, individuals in the fittest class evolve approximately neutrally. Fluctuations in the size of this class propagate to the mean, which in turn generates a delayed restoring force opposing the fluctuation. We have shown that the variance of the fluctuations in the population $\n_0$ of the top bin is proportional to $\n_0/s$ and increases as $\log \lambda$ with the ratio $\lambda$ of the mutation rate $\mut$ and the mutational effect $s$. Fluctuations of $\n_0$ perturb the mean after a time $\sim s^{-1}\log \lambda$. These two observations have a straightforward connection: sampling fluctuations can accumulate without a restoring force for a time $s^{-1} \log \lambda$. During this time, the typical perturbation of the top bin by drift is $\sim \sqrt{\n_0 s^{-1}\log \lambda}$ and hence the variance is $\approx  \n_0 s^{-1}\log \lambda$. We have used these insights into the coupling between $\n_0$, the mean fitness, and the resulting delayed restoring force on fluctuations of $\n_0$ to approximate the rate of Muller's ratchet. 

The history dependence of the restoring force has not been accounted for in previous analysis of the rate of Muller's ratchet \citep{Haigh:1978p37141,Stephan:1993p42929,Gordo:2000p42688,Jain:2008p45047} who introduced a constant factor to parameterize the effective strength of the selection opposing fluctuations in the top bin, or \citet{Waxman:2010p47020}, who replaced all mutant classes by one effective class and thereby mapped the problem to the fixation of a deleterious allele. We have shown that  to achieve agreement between theory and numerical simulation one must account for the delayed nature of selection acting on fluctuations. Comparing our final expression for the ratchet rate with that given previously \citep{Stephan:1993p42929,Gordo:2000p42688,Jain:2008p45047}, the history dependence manifests itself as a decreasing effective strength of selection with increasing $\lambda =\mut/s$. This decrease is due to a larger temporal delay of the response of the mean fitness to fluctuations of the least loaded class. History dependence is a general consequence of projecting a multi-dimensional stochastic dynamics onto a lower dimensional space (here, the size $\n_0$ of the fittest class). Such memory effects can be accounted for by the path-integral formulation of stochastic processes which we used to approximate the rate of Muller's ratchet.

Even though the model is extremely simplistic and the sensitive dependence of the ratchet rate on poorly known parameters such as the effect size of mutations, population size, and mutation rate, precludes quantitative comparison with the real world, we believe that some general lessons can be learned from our analysis. The propagation of fluctuations from the fittest to less fit individuals is expected to be a generic feature of many models and natural populations. 
In particular, very similar phenomena arise in the dynamics of adapting populations driven by the accumulation of beneficial mutations \citep{Tsimring:1996p19688,Rouzine:2003p33590,Rouzine:2008p20864,Cohen:2005p45154,Desai:2007p954,Neher:2010p30641,Hallatschek:2011p39697}. The speed of these traveling waves is typically determined by stochastic effects at the high fitness edges. We expect that the fluctuations of the speed of adaptation can be understood and quantified with the concepts and tools that we introduced above.

Populations spread out in fitness  have rather different coalescence properties than neutral populations, which are described by Kingman's coalescent \citep{Kingman:1982p28911}. These differences go beyond the familiar reduction in effective population size and distortions of genealogies due to background selection \citep{Charlesworth:1993p36005,Higgs:1995p45226,Walczak:2011p45228}. The most recent common ancestor of such populations most likely derives from this high fitness tail and fluctuations of this tail determine the rate at which lineages merge and thereby the genetic diversity of the population \citep{Brunet:2007p18866,Rouzine:2007p17401,Neher:2011p42539}. Thus, quantitative understanding of fluctuations of fitness distributions is also essential for understanding non-neutral coalescent processes.


Generalizing the analysis of fluctuations of fitness distributions to adapting ``traveling waves" and the study of their implications for the coalescent properties of the population are interesting avenues for future research.

\section{Acknowledgements}
We are grateful for stimulating discussions with Michael Desai, Dan Balick and Sid Goyal.  RAN is supported by an ERC-starting grant HIVEVO 260686 and BIS acknowledges support from NIH under grant GM086793. This research was also supported in part by the NSF under Grant No.~NSF PHY11-25915.

\bibliography{bib}

\begin{thebibliography}{34}

\bibitem[{\sc Brunet {\em et~al.\/}}(2007){\sc Brunet, Derrida, Mueller, {\rm
  and} Munier}]{Brunet:2007p18866}
{\sc Brunet, E., B.~Derrida, A.~H. Mueller, {\rm and} S.~Munier}, 2007 Effect
  of selection on ancestry: an exactly soluble case and its phenomenological
  generalization. Physical review E, Statistical, nonlinear, and soft matter
  physics {\bf 76}: 041104.

\bibitem[{\sc Charlesworth}(2012)]{Charlesworth:2012p45100}
{\sc Charlesworth, B.}, 2012 The effects of deleterious mutations on evolution
  at linked sites. Genetics {\bf 190}: 5--22.

\bibitem[{\sc Charlesworth {\em et~al.\/}}(1993){\sc Charlesworth, Morgan, {\rm
  and} Charlesworth}]{Charlesworth:1993p36005}
{\sc Charlesworth, B., M.~T. Morgan, {\rm and} D.~Charlesworth}, 1993 The
  effect of deleterious mutations on neutral molecular variation. Genetics {\bf
  134}: 1289--303.

\bibitem[{\sc Cohen {\em et~al.\/}}(2005){\sc Cohen, Kessler, {\rm and}
  Levine}]{Cohen:2005p45154}
{\sc Cohen, E., D.~A. Kessler, {\rm and} H.~Levine}, 2005 Front propagation up
  a reaction rate gradient. Phys Rev E Stat Nonlin Soft Matter Phys {\bf 72}:
  066126.

\bibitem[{\sc Desai {\rm and} Fisher}(2007)]{Desai:2007p954}
{\sc Desai, M.~M. {\rm and} D.~S. Fisher}, 2007 Beneficial mutation selection
  balance and the effect of linkage on positive selection. Genetics {\bf 176}:
  1759--98.

\bibitem[{\sc Etheridge {\em et~al.\/}}(2007){\sc Etheridge, Pfaffelhuber, {\rm
  and} Wakolbinger}]{Etheridge:2007p44291}
{\sc Etheridge, A., P.~Pfaffelhuber, {\rm and} A.~Wakolbinger}, 2007 How often
  does the ratchet click? facts, heuristics, asymptotics. In {\em Trends in
  Stochastic Analysis\/}, edited by P.~M. Jochen~Blath {\rm and} M.~Scheutzow,
  pp. 365--390, Cambridge University Press 2009.

\bibitem[{\sc Felsenstein}(1974)]{Felsenstein:1974p23937}
{\sc Felsenstein, J.}, 1974 The evolutionary advantage of recombination.
  Genetics {\bf 78}: 737--56.

\bibitem[{\sc Feynman {\rm and} Hibbs}(1965)]{Feynman:1965}
{\sc Feynman, R.~P. {\rm and} A.~R. Hibbs}, 1965 {\em Quantum mechanics and
  Path Integrals\/}. McGraw-Hill Inc, New York.

\bibitem[{\sc Gardiner}(2004)]{WGardiner:2004p36981}
{\sc Gardiner, C.~W.}, 2004 {\em Handbook of stochastic methods for Physics,
  Chemistry and the Natural sciences\/}. Springer.

\bibitem[{\sc Gessler}(1995)]{Gessler:1995p42788}
{\sc Gessler, D.~D.}, 1995 The constraints of finite size in asexual
  populations and the rate of the ratchet. Genet Res {\bf 66}: 241--53.

\bibitem[{\sc Gordo {\rm and} Charlesworth}(2000)]{Gordo:2000p42688}
{\sc Gordo, I. {\rm and} B.~Charlesworth}, 2000 The degeneration of asexual
  haploid populations and the speed of {Muller's} ratchet. Genetics {\bf 154}:
  1379--87.

\bibitem[{\sc Goyal {\em et~al.\/}}(2012){\sc Goyal, Balick, Jerison, Neher,
  Shraiman, {\rm and} Desai}]{Goyal:2012p47382}
{\sc Goyal, S., D.~J. Balick, E.~R. Jerison, R.~A. Neher, B.~I. Shraiman, {\rm
  and} M.~M. Desai}, 2012 Dynamic mutation selection balance as an evolutionary
  attractor. Genetics .

\bibitem[{\sc Gradshteyn {\rm and} Ryzhik}(2007)]{GR}
{\sc Gradshteyn, I.~S. {\rm and} I.~M. Ryzhik}, 2007 {\em Table of
  {{I}}ntegrals, {{S}}eries, and {{P}}roducts\/}. Academic Press, New York.

\bibitem[{\sc Haigh}(1978)]{Haigh:1978p37141}
{\sc Haigh, J.}, 1978 The accumulation of deleterious genes in a population --
  {Muller}'s ratchet. Theoretical Population Biology {\bf 14}: 251--67.

\bibitem[{\sc Hallatschek}(2011)]{Hallatschek:2011p39697}
{\sc Hallatschek, O.}, 2011 The noisy edge of traveling waves. Proceedings of
  the National Academy of Sciences of the United States of America {\bf 108}:
  1783--7.

\bibitem[{\sc Higgs {\rm and} Woodcock}(1995)]{Higgs:1995p45226}
{\sc Higgs, P. {\rm and} G.~Woodcock}, 1995 The accumulation of mutations in
  asexual populations and the structure of genealogical trees in the presence
  of selection. J. Math. Biol. {\bf 33}: 677--102.

\bibitem[{\sc Jain}(2008)]{Jain:2008p45047}
{\sc Jain, K.}, 2008 Loss of least-loaded class in asexual populations due to
  drift and epistasis. Genetics {\bf 179}: 2125--34.

\bibitem[{\sc Kingman}(1982)]{Kingman:1982p28911}
{\sc Kingman, J.}, 1982 On the genealogy of large populations. Journal of
  Applied Probability {\bf 19 IS -}: 27--43.

\bibitem[{\sc Lau {\rm and} Lubensky}(2007)]{Lau:2007p45316}
{\sc Lau, A. W.~C. {\rm and} T.~C. Lubensky}, 2007 State-dependent diffusion:
  Thermodynamic consistency and its path integral formulation. Phys Rev E Stat
  Nonlin Soft Matter Phys {\bf 76}: 011123.

\bibitem[{\sc Lynch {\em et~al.\/}}(1993){\sc Lynch, B{\"u}rger, Butcher, {\rm
  and} Gabriel}]{Lynch:1993p42844}
{\sc Lynch, M., R.~B{\"u}rger, D.~Butcher, {\rm and} W.~Gabriel}, 1993 The
  mutational meltdown in asexual populations. J Hered {\bf 84}: 339--44.

\bibitem[{\sc Muller}(1964)]{Muller:1964p45018}
{\sc Muller, H.~J.}, 1964 The relation of recombination to mutational advance.
  Mutat Res {\bf 106}: 2--9.

\bibitem[{\sc Neher {\rm and} Shraiman}(2011)]{Neher:2011p42539}
{\sc Neher, R.~A. {\rm and} B.~I. Shraiman}, 2011 Genetic draft and
  quasi-neutrality in large facultatively sexual populations. Genetics {\bf
  188}: 975--996.

\bibitem[{\sc Neher {\em et~al.\/}}(2010){\sc Neher, Shraiman, {\rm and}
  Fisher}]{Neher:2010p30641}
{\sc Neher, R.~A., B.~I. Shraiman, {\rm and} D.~S. Fisher}, 2010 Rate of
  adaptation in large sexual populations. Genetics {\bf 184}: 467--481.

\bibitem[{\sc Oliphant}(2007)]{Oliphant:2007p25672}
{\sc Oliphant, T.}, 2007 Python for scientific computing. Computing in Science
  {\&} Engineering {\bf 9}: 10--20.

\bibitem[{\sc Pfaffelhuber {\em et~al.\/}}(2012){\sc Pfaffelhuber, Staab, {\rm
  and} Wakolbinger}]{Pfaffelhuber:2011p44301}
{\sc Pfaffelhuber, P., P.~R. Staab, {\rm and} A.~Wakolbinger}, 2012 Muller's
  ratchet with compensatory mutations. to appear in Annals of Applied
  Probability {\bf xxx}, 26 pages, 3 figures.

\bibitem[{\sc Rice}(1987)]{Rice:1987p45218}
{\sc Rice, W.~R.}, 1987 Genetic hitchhiking and the evolution of reduced
  genetic activity of the {Y} sex chromosome. Genetics {\bf 116}: 161--7.

\bibitem[{\sc Rouzine {\em et~al.\/}}(2008){\sc Rouzine, Brunet, {\rm and}
  Wilke}]{Rouzine:2008p20864}
{\sc Rouzine, I.~M., E.~Brunet, {\rm and} C.~O. Wilke}, 2008 The traveling-wave
  approach to asexual evolution: Muller's ratchet and speed of adaptation.
  Theoretical Population Biology {\bf 73}: 24--46.

\bibitem[{\sc Rouzine {\rm and} Coffin}(2007)]{Rouzine:2007p17401}
{\sc Rouzine, I.~M. {\rm and} J.~M. Coffin}, 2007 Highly fit ancestors of a
  partly sexual haploid population. Theoretical Population Biology {\bf 71}:
  239--50.

\bibitem[{\sc Rouzine {\em et~al.\/}}(2003){\sc Rouzine, Wakeley, {\rm and}
  Coffin}]{Rouzine:2003p33590}
{\sc Rouzine, I.~M., J.~Wakeley, {\rm and} J.~M. Coffin}, 2003 The solitary
  wave of asexual evolution. Proc Natl Acad Sci USA {\bf 100}: 587--92.

\bibitem[{\sc Stephan {\em et~al.\/}}(1993){\sc Stephan, Chao, {\rm and}
  Smale}]{Stephan:1993p42929}
{\sc Stephan, W., L.~Chao, {\rm and} J.~G. Smale}, 1993 The advance of
  {Muller's} ratchet in a haploid asexual population: approximate solutions
  based on diffusion theory. Genet Res {\bf 61}: 225--31.

\bibitem[{\sc Stephan {\rm and} Kim}(2002)]{Stephan:2002_review}
{\sc Stephan, W. {\rm and} Y.~Kim}, 2002 Recent applications of diffusion
  theory to population genetics. In {\em Modern Developments in Theoretical
  Population Genetics\/}, edited by M.~Slatkin {\rm and} M.~Veuille, pp.
  72--93, Oxford University Press, Oxford, UK.

\bibitem[{\sc Tsimring {\em et~al.\/}}(1996){\sc Tsimring, Levine, {\rm and}
  Kessler}]{Tsimring:1996p19688}
{\sc Tsimring, L., H.~Levine, {\rm and} D.~Kessler}, 1996 {RNA} virus evolution
  via a fitness-space model. Phys Rev Lett {\bf 76}: 4440--4443.

\bibitem[{\sc Walczak {\em et~al.\/}}(2011){\sc Walczak, Nicolaisen, Plotkin,
  {\rm and} Desai}]{Walczak:2011p45228}
{\sc Walczak, A.~M., L.~E. Nicolaisen, J.~B. Plotkin, {\rm and} M.~M. Desai},
  2011 The structure of genealogies in the presence of purifying selection: A
  "fitness-class coalescent". Genetics .

\bibitem[{\sc Waxman {\rm and} Loewe}(2010)]{Waxman:2010p47020}
{\sc Waxman, D. {\rm and} L.~Loewe}, 2010 A stochastic model for a single click
  of muller's ratchet. Journal of Theoretical Biology {\bf 264}: 1120--1132.

\end{thebibliography}

\newpage
\appendix
\section{Supplement}

To analyze the covariances of different fitness classes, we begin with Eq.~(7) of the main text, which expresses $\dx_k(\tau)$ in terms of eigenvectors $\dx_k(\tau) = \sum_j \mr{j}_k a_j(\tau)$. Projecting on the left eigenvectors then results in equations for $a_j(\tau)$
\begin{equation}
\label{eq:app_modes}
\frac{d}{d\tau} a_j(\tau) = -j a_j(\tau) + \sum_k \ml{j}_k \sqrt{\frac{\xs_k}{Ns}} \eta_k(\tau)
\end{equation}
where $\eta_k(\tau)$ are uncorrelated Gaussian white noise terms with $\la \eta_k(\tau) \eta_l(\tau')\ra = \delta_{kl}\delta(\tau-\tau')$. Since each noise term $\eta_k$ contributes to all $a_j$, the noise induces correlated fluctuations of the $a_j(\tau)$, which we need to understand in order to analyze the fluctuations of the fitness distributions. The inhomogeneous \EQ{app_modes} has the  solution
\begin{equation}
a_j(\tau) = \int_{-\infty}^{\tau}d\tau' e^{-j(\tau-\tau')} \sum_k \ml{j}_k \sqrt{\frac{\xs_k}{Ns}} \eta_k(\tau')
\end{equation}
The autocorrelation function of the loadings of different eigendirections separated by $\Delta \tau$ in time is therefore given by
\begin{equation}
\begin{split}
\langle a_i(\tau) a_j(\tau+\Delta\tau)\rangle &= \int_{-\infty}^\tau d\tau'\int_{-\infty}^{\tau+\Delta\tau} d\tau'' e^{-i(\tau-\tau')-j(\tau+\Delta\tau -\tau'')}\sum_{k,l} \frac{\ml{i}_k\ml{j}_l\sqrt{\xs_k\xs_l}}{Ns} \langle \eta_k(\tau')\eta_l(\tau'')\rangle\\
&=  \int_{-\infty}^\tau d\tau' e^{-i(\tau-\tau')-j(\tau+\Delta\tau -\tau')}\sum_{k} \frac{\ml{i}_k\ml{j}_k\xs_k}{Ns}\\
&= \frac{e^{-j\Delta\tau}}{i+j}\sum_{k} \frac{\ml{i}_k\ml{j}_k\xs_k}{Ns}
\end{split}
\end{equation}
where we have used $\la \eta_k(\tau) \eta_l(\tau')\ra = \delta_{kl}\delta(\tau-\tau')$.

\subsection*{Correlation functions $n_0$ and the mean}
To calculate the variances and covariance of $\x_0$ and the mean fitness, we express them in terms of the eigenmodes $a_j(\tau)$ ($j>0$)
\begin{eqnarray}
\dx_0(\tau) &=& \sum_{j>0} \mr{j}_0 a_j(\tau)  = -e^{-\lambda} \sum_{j>0} a_j(\tau)\\
\dk(\tau) &=& -\sum_{j>0,k} k\mr{j}_k a_j(\tau) = -\sum_{j>0,k} k(\xs_{k-j}-\xs_k) a_j(\tau) =  -\sum_{j>0} j a_j(\tau)
\end{eqnarray}

\subsection*{The auto-correlation of $\x_0$}
The auto-correlation of $\x_0$ is given by
\begin{equation}
\label{eq:n0var}
\begin{split}
\la \x_0(\tau) \x_0(\tau+\Delta\tau)\ra &=  e^{-2\lambda} \sum_{i,j>0} \frac{e^{-j\Delta\tau}}{i+j}\sum_{k} \frac{\ml{i}_k\ml{j}_k\xs_k}{Ns} \\
& = \frac{e^{-\lambda}}{Ns} \sum_{i,j>0} \frac{\lambda^{i+j}e^{-j\Delta\tau}}{i+j}\sum_{k=0}^{\min(i,j)} \frac{(-1)^{i+j}\lambda^{-k}}{(j-k)!(i-k)!k!} \\
& = \frac{e^{-\lambda}}{Ns} \int_0^\infty dz \sum_{i,j>0} e^{-z(i+j)}\lambda^{i+j}e^{-j\Delta\tau}\sum_{k=0}^{\min(i,j)} \frac{(-1)^{i+j}\lambda^{-k}}{(j-k)!(i-k)!k!} 
\end{split}
\end{equation}
Let us focus on the triple sum inside the integral and simplify it by introducing $a=-\lambda e^{-z}$ and $b=-\lambda e^{-z-\Delta\tau}$. Furthermore, let us look at the $i=j$ and the $i\neq j$ contributions separately. The diagonal contribution  ($i=j$) is
\begin{equation}
\begin{split}
&\sum_{i=0} a^ib^i \sum_{k=0}^{i} \frac{\lambda^{-k}}{(i-k)!(i-k)!k!} = \sum_{i>0} \sum_{k=0}^{i} \frac{(ab)^{i-k}(ab)^{k}\lambda^{-k}}{k!((i-k)!)^2}\\
&=\sum_{k>0} \sum_{i\geq k} \frac{(ab)^{i-k}(ab)^{k}\lambda^{-k}}{k!((i-k)!)^2} + \sum_{i>0} \frac{(ab)^{i}}{(i!)^2} \\
&=\sum_{k>0} \frac{(ab)^{k}\lambda^{-k}}{k!}\sum_{n\geq 0} \frac{(ab)^{n}}{(n!)^2} + J_0(-\cp 2\sqrt{ab})-1   \quad\quad \mathrm{using}\quad n=i-k\\
&=(e^{ab/\lambda}-1)J_0(-\cp 2\sqrt{ab}) + J_0(-\cp 2\sqrt{ab})-1\\
&=e^{ab/\lambda} J_0(-\cp 2\sqrt{ab})-1
\end{split}
\end{equation}
where $J_n(z)$ is the $n$th Bessel function of first kind, and $\cp=\sqrt{-1}$. 
When evaluating the off-diagonal contribution, we will encounter terms like
\begin{equation}
\sum_{k>0} \frac{(ab)^{k}}{k!(k+m)!} = \sum_{k} \frac{(ab)^{k}}{k!(k+m)!} - \frac{1}{m!} = \frac{J_m(2\cp\sqrt{ab})}{(\cp\sqrt{ab})^m}-\frac{1}{m!}
\end{equation}
The off-diagonal contribution can be further split into the parts $i>j$ and $i<j$ which can be evaluated as follows:
\begin{equation}
\begin{split}
&\sum_{0<i<j} a^{i}b^j\sum_{k\geq 0}^{i} \frac{\lambda^{-k}}{k!(j-k)!(i-k)!} 
= \sum_{i>0}\sum_{j>i} a^{i}b^{j-i+i}\sum_{k\geq 0}^{i} \frac{\lambda^{-k}}{k!(i+(j-i)-k)!(i-k)!}\\
& = \sum_{i>0}\sum_{m>0} a^{i}b^{m+i}\sum_{k\geq 0}^{i} \frac{\lambda^{-k}}{k!(i+m-k)!(i-k)!}  \quad\quad \mathrm{using}\quad m=j-i \\
&=\sum_{k>0}\sum_{m>0} \sum_{i\geq k} \frac{a^{i}b^{m+i}\lambda^{-k}}{k!(i+m-k)!(i-k)!} + \sum_{m>0} \sum_{i>0} \frac{a^{i}b^{m+i}}{(i+m)! i!} \\
&=\sum_{k>0}\sum_{m>0}\sum_{n\geq 0} \frac{ a^{n+k}b^{m+n+k}\lambda^{-k}}{k!(n+m)!n!} + \sum_{m>0} b^{m}\sum_{i>0} \frac{(ab)^{i}}{(i+m)! i!}  \quad\quad \mathrm{using}\quad n=i-k\\
&=\sum_{k>0}\sum_{m>0}\frac{ a^{k}b^{m+k}\lambda^{-k}}{k!}\frac{J_m(2\cp\sqrt{ab})}{(\cp\sqrt{ab})^m} + \sum_{m>0} b^{m}\left(\frac{J_m(2\cp\sqrt{ab})}{(\cp\sqrt{ab})^m} -\frac{1}{m!}\right)\\
&=\sum_{m>0}\sum_{k\geq 0}\frac{  a^{k}b^{k}b^{m/2}a^{-m/2}\lambda^{-k}}{k!}\frac{J_m(2\cp\sqrt{ab})}{(\cp)^m} -e^b+1 = \sum_{m>0} \left(\frac{b}{a}\right)^{m/2}e^{ab/\lambda}\frac{J_m(2\cp\sqrt{ab})}{(-1)^{m/2}} -e^b+1\\ &=e^{ab/\lambda}\sum_{m>0} \left(\frac{b}{a}\right)^{m/2}(-\cp)^m J_m(2\cp\sqrt{ab}) -e^b+1
\end{split}
\end{equation}
The off-diagonal terms for $i>j$ is obtained by interchanging $a$ and $b$ such that the full off-diagonal contribution is 
\begin{equation}
e^{ab/\lambda}\sum_{m>0} \left[\left(\frac{b}{a}\right)^{m/2}+\left(\frac{a}{b}\right)^{m/2}\right](-\cp)^m J_m(2\cp\sqrt{ab}) -e^a-e^b+2
\end{equation}
Next, we use the definition of the generating function of the Bessel functions (\citet{GR}, 8.511)
\begin{equation}
e^{\frac{1}{2}(t-t^{-1})z} =J_0(z) +  \sum_{m>0} (t^m+(-t)^{-m}) J_m(z) 
\end{equation}
which turns the off-diagonal contribution into 
\begin{equation}
e^{ab/\lambda}\left(e^{\sqrt{ab} \left(\sqrt{\frac{b}{a}}+\sqrt{\frac{a}{b}}\right)}-J_0(2\cp\sqrt{ab})\right)-e^a-e^b+2
\end{equation}
Combining the diagonal and off-diagonal contributions and substituting $a$ and $b$, we find for the integrand in \EQ{n0var}
\begin{equation}
e^{ab/\lambda+\sqrt{ab} \left(\sqrt{\frac{b}{a}}+\sqrt{\frac{a}{b}}\right)}-e^a-e^b+1 = 
e^{\lambda e^{-2z-\Delta \tau}-\lambda e^{-z} -\lambda e^{-z-\Delta \tau}}-e^{-\lambda e^{-z}}-e^{-\lambda e^{-z-\Delta \tau}}+1
\end{equation}
The auto-correlation of $\x_0$ is therefore given by
\begin{equation}
\begin{split}
\la \x_0(\tau) \x_0(\tau+\Delta\tau)\ra & = \frac{e^{-\lambda}}{Ns} \int_0^\infty dz \left(e^{\lambda e^{-2z-\Delta\tau}}e^{-\lambda(e^{-z}+e^{-z-\Delta\tau})}-e^{-\lambda e^{-z}}-e^{-\lambda e^{-z-\Delta\tau}}+1\right) \\
&= \frac{e^{-\lambda}}{Ns} \int_0^1 \frac{d\theta}{\theta} \left(e^{\lambda \theta^2 e^{-\Delta\tau} -\lambda\theta(1+e^{-\Delta\tau})}-e^{-\lambda \theta}-e^{-\lambda \theta e^{-\Delta\tau}}+1\right)
\end{split}
\end{equation}

\subsection*{Auto-correlation of the mean fitness}
The autocorrelation function of the mean is defined as
\begin{equation}
\begin{split}
\la \dk(\tau) \dk(\tau+\Delta\tau) \ra &= \sum_{i,j>0} ij \la a_i(\tau)a_j(\tau+\Delta\tau)\ra \\
&= \partial_\mu \partial_\nu \frac{1}{Ns} \int_0^\infty dz \sum_{i,j>0} \mu^i \nu^j e^{-z(i+j)}\lambda^{i+j}e^{-j\Delta\tau}\sum_{k=0}^{\min(i,j)} \frac{(-1)^{i+j}\lambda^{-k}}{(j-k)!(i-k)!k!}
\end{split}
\end{equation}
where the last line is to be evaluated at $\nu=\mu=1$. Hence the problem is reduced to the one already solved with $a=-\mu\lambda e^{-z}$ and $b=-\nu\lambda e^{-z-\Delta\tau}$. We find 
\begin{equation}
\begin{split}
\langle\dk(\tau) \dk(\tau+\Delta\tau )\rangle  
 & =\frac{\lambda e^{\lambda}}{Ns}\int_0^1 d\theta e^{-\Delta\tau } e^{\lambda \theta^2e^{-\Delta\tau } -\lambda(1+e^{-\Delta\tau }) \theta} \left(\theta  +\lambda \theta\left(\theta e^{-\Delta\tau }- 1\right)\left(\theta- 1\right)\right)
 \end{split}
\end{equation}

\subsection*{Cross-correlation of $\x_0$ and the mean fitness}
When calculating the cross-correlation between $\x_0$ and the mean fitness we have to distinguish the cases where $\x_0$ precedes the mean fitness and vice-versa. Otherwise, the calculation proceeds almost unchanged from the cases discussed above.
\begin{equation}
\begin{split}
\la \x_0(\tau) \dk(\tau+\Delta\tau) \ra &= \sum_{i,j>0} j \la a_i(\tau)a_j(\tau+\Delta\tau)\ra \\
&= \partial_\nu \frac{1}{Ns} \int_0^\infty dz \sum_{i,j>0} a^i b^j\sum_{k=0}^{\min(i,j)} \frac{(-1)^{i+j}\lambda^{-k}}{(j-k)!(i-k)!k!}
\end{split}
\end{equation}
with $a=-\lambda e^{-z}$, $b=-\nu \lambda e^{-z-\Delta\tau}$ if $\Delta\tau>0$ and $a=-\lambda e^{-z+\Delta\tau}$, $b=-\nu \lambda e^{-z}$ if $\Delta\tau<0$. The result is 
\begin{equation}
\langle \dx_0(\tau) \dk(\tau+\Delta\tau)\rangle   = \frac{\lambda}{Ns} \begin{cases} 
  e^{-\Delta\tau} \int_0^1 d\theta \left((\theta-1)e^{e^{-\Delta\tau}\lambda \theta^2-\lambda(1+e^{-\Delta\tau}) \theta}+ e^{-e^{-\Delta\tau}\lambda \theta}\right) & \Delta\tau >0 \\
 \int_0^1 d\theta \left((e^{\Delta\tau} \theta-1)e^{e^{\Delta\tau}\lambda \theta^2-\lambda(1+e^{\Delta\tau}) \theta}+e^{-\lambda \theta}\right) & \Delta\tau <0 
\end{cases}
\end{equation}

\begin{figure}[htp]
\begin{center}
  \includegraphics[width=0.7\columnwidth]{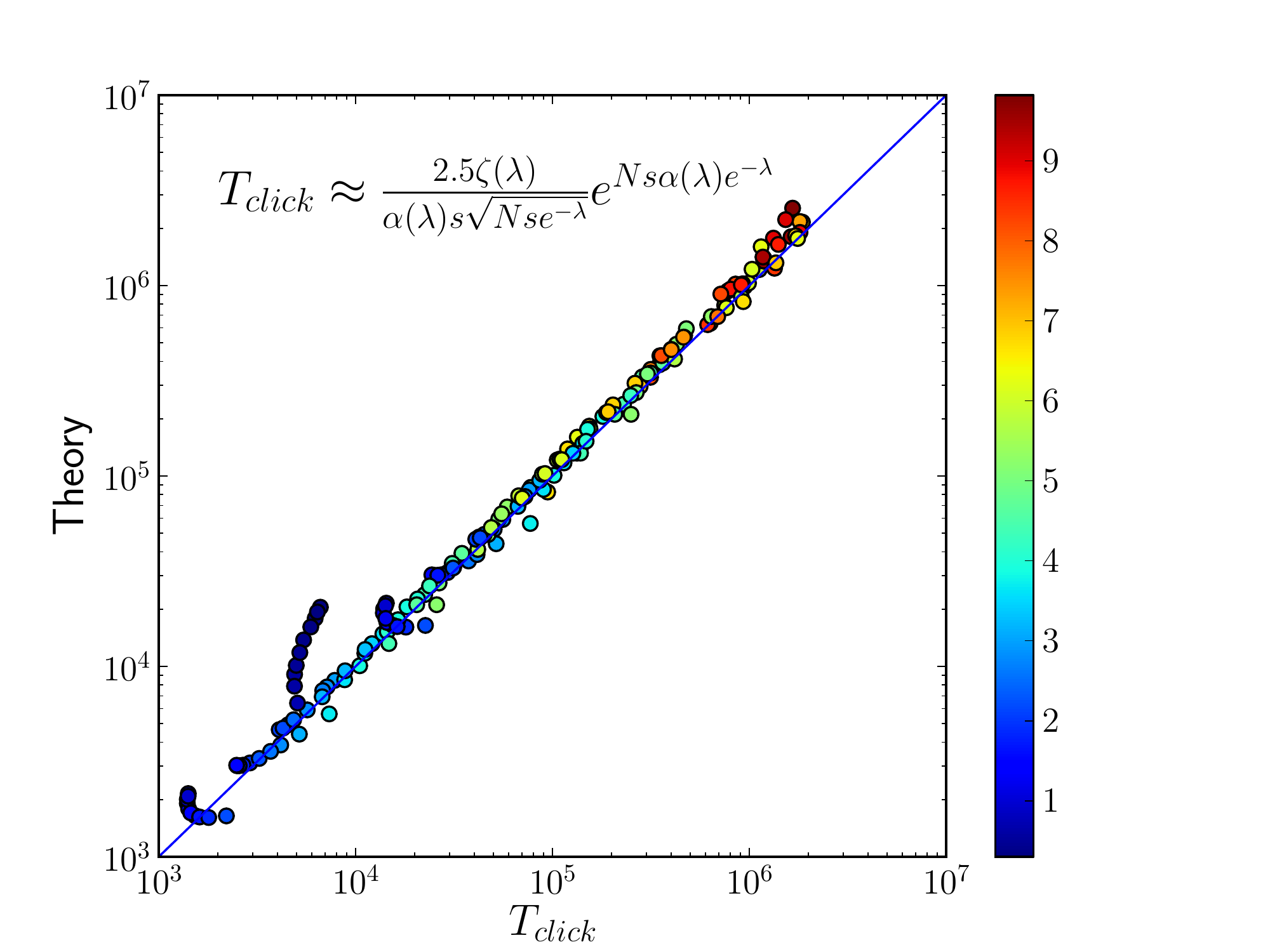}
  \caption[labelInTOC]{The approximation of the mean time between clicks of the ratchet is accurate over a large range of parameters if $Ns\alpha(\lambda)e^{-\lambda}$ is large compared to one. $Ns\alpha(\lambda)e^{-\lambda}$ determines whether the clicks of the ratchet are far apart compared to the relaxation time of the distribution and is indicated as the color of the data points. The condition  $Ns\alpha(\lambda)e^{-\lambda}>1$ is violated for the fastest clicks shown, resulting in the deviation of the dark blue points.}
  \label{fig:explicit}
\end{center}
\end{figure}

\end{document}